\documentstyle[eqsecnum,preprint,aps]{revtex}
\newcommand{\be}{\begin{equation}}
\newcommand{\ee}{\end{equation}}
\newcommand{\bea}{\begin{eqnarray}}
\newcommand{\eea}{\end{eqnarray}}
\newcommand{\OP}{\vec{\psi}}
\newcommand{\SG}{\vec{\sigma}}
\newcommand{\M}{\vec{m}}
\newcommand{\U}{\vec{u}}
\newcommand{\F}{{\cal F}}

\begin{document}
\draft
\title{Vortex annihilation \\ in the ordering
kinetics of the $O(2)$ model}
\author{Gene F. Mazenko and Robert A. Wickham}
\address{The James Franck Institute and the Department of Physics\\ The
University of Chicago\\ Chicago, Illinois 60637}
\date{\today}
\maketitle
%
%  ABSTRACT
%
\begin{abstract}
The vortex-vortex and vortex-antivortex correlation functions are
determined for the two-dimensional $O(2)$ model undergoing phase ordering.
We find reasonably good agreement
with simulation results for the vortex-vortex
correlation function where there is a short-scaled distance depletion
zone due to the repulsion of like-signed vortices.  The
vortex-antivortex correlation function agrees well with simulation
results for intermediate and long-scaled distances.
At short-scaled   distances the simulations show a depletion zone not seen in
the theory.
\end{abstract}
\draft
%
% INTRODUCTION
%
\section{Introduction}

The late-stage ordering of systems in the process of breaking a
continuous symmetry is dominated by the dynamics of topological
defects.  In the case of the $n$-vector model with the number
of components  $n$ of the order parameter equal to the spatial dimensionality
$d$ one has point defects which are vortices for $n=2$ and monopoles for $n=3$
\cite{MERMIN79}. We focus mainly on the case $n=d=2$ here.
Because of the conservation of topological charge, the
ordering in these systems is through the charge conserving process of
vortex-antivortex annihilation.  The statistical description
of this annihilation process will be a central focus of this paper.
We present here the first calculation for the separate
vortex-vortex and vortex-antivortex correlation functions. Knowledge of these
functions is an important ingredient in understanding the vortex annihilation
process. Since these functions contain detailed information about the
vortex correlations, comparison of our results with simulation and experiment
provides the most stringent test of the theory to date.
In the original work \cite{LIU92b} in this area only the signed defect
correlation function was computed. The unsigned quantity is technically much
more difficult to evaluate as we discuss in this paper. The comparison of the
result for the signed quantity with simulations was confused by the use of a
theory \cite{LIU92a}
for describing order- parameter fluctuations which led to unphysical
singularities at short distances in this quantity. The source of this
singularity was recently uncovered by us \cite{MAZENKO96} and the theory
reorganized so as to give physical results for the signed defect
correlation function. This development motivated us to make a renewed effort
to evaluate the unsigned defect correlation function, with the results
presented here.

We show in Figure \ref{FIG:CVV}
the vortex-vortex correlation function for the
two-dimensional $O(2)$ model for simulations \cite{MONDELLO90} and the
theory developed here. Beautiful experiments \cite{NAGAYA95} on
two-dimensionally aligned nematic liquid crystals, which mimic the $O(2)$
model, have also been able to measure the vortex correlations.
Their results essentially track the simulation results, but are more noisy, so
we do not show them here. There is reasonable agreement between the theory
and the simulation results. Both show a depletion zone at short-scaled
distances for like-signed defects.  This is expected on physical grounds
since like-signed defects repel one another.  It is non-trivial
that we see evidence for a repulsive interaction. The vortex interactions
emerge naturally in the theory from the equation of motion for the
order-parameter.
In Figure \ref{FIG:CVA} we show the
vortex-antivortex correlation function for
theory and simulation \cite{MONDELLO90}. There is
good agreement at large and intermediate scaled distances.
There is a clear discrepancy between theory and simulation results
at short-scaled distances.  The theory shows a monotonic
behavior as the separation distance goes to zero.  The simulation,
however, shows a maximum at short separation distances and then
falls rapidly to zero.  The depletion zone seen in
the simulation data for the vortex-antivortex correlation function is harder
to understand physically since the pair is attractive and headed toward
annihilation.
Some possible explanations will be presented in Section VIII.
While the theory satisfies the sum rule implied by topological charge
conservation, it does not appear that this general constraint is satisfied
by the simulations.
%
% MODEL
%
\section{Model}

We consider the $n$-vector model with $O(n)$ symmetry, which describes the
dynamics of a  non-conserved, $n$-component order-parameter field $\OP (1) =
(\psi_{1} (1), \cdots,\psi_{n} (1) )$. Here we use the shorthand notation
$1 = ({\bf r}_{1},t_{1})$. The order-parameter evolves {\em via} the
time-dependent Ginzburg-Landau equation \cite{LIU92a}
%
% EQUATION OF MOTION
\be
\frac{\partial \OP}{\partial t} = \nabla^{2} \OP -
\frac{\partial V[\OP]}{\partial \OP}
\label{EQ:MOT}
\ee
which can be derived from a free-energy containing a square-gradient term and a
potential term, $V[\OP]$. We assume that the quench is to zero temperature
where the usual noise term on the right-hand side of (\ref{EQ:MOT})
is zero \cite{BRAY89}. The potential  $V[\OP]$ is chosen to have
$O(n)$ symmetry with a degenerate set of equilibrium minima at
$\psi \equiv |\OP| = \psi_{0}$. Since only these properties
of $V[\OP]$ will be important in what follows we need not be more
specific in our choice for $V[\OP]$.
It is believed that our final results are independent of the exact nature
of the initial state, provided it is a disordered state.
Indeed, it is well-established \cite{BRAYREV}
that for late times $t$ following a quench from the disordered to the
ordered phase the dynamics obey scaling and the system can be described in
terms of a single growing length $L(t)$, which is characteristic of the
spacing between defects. In this scaling regime the order-parameter
correlation function
%
% ORDER PARAMETER CORRELATION FUNCTION
%
\be
\label{EQ:OPCOR}
C(12) \equiv \langle \OP (1) \cdot \OP (2) \rangle
\ee
has an equal-time scaling form
%
% SCALING FORM
%
\be
\label{EQ:OPSCAL}
C({\bf r},t) = \psi_{0}^{2} \F(x),
\ee
where $\F$ is a universal function of $x = r/L(t)$
($r \equiv |{\bf r}| \equiv |{\bf r_{2}} - {\bf r_{1}}|$), depending only
on $n$ and $d$. It is also well-established that, in the scaling regime,
$L(t) \sim t^{\phi}$. For the non-conserved models considered here the
exponent $\phi = 1/2$ \cite{N=2D=2}.
%
% GENERAL FEATURES OF POINT DEFECTS
%
\section{General Features of Point Defects}

Let us briefly review the theoretical picture associated
with topological point defects. The results of this section are quite general
and hold independent of the particular dynamics, such as (\ref{EQ:MOT}), or the
approximation scheme, such as the gaussian approximation, used to model the
system. The vortex charge density can be written in the form
%
% VORTEX DENSITY
%
\be
\rho(1) = \sum_{\alpha} q_{\alpha} \mbox{ }
          \delta({\bf r_{1}} - {\bf x_{\alpha}}(t_{1}))
\ee
where ${\bf x_{\alpha}}(t_{1})$ is the position at time $t_{1}$ of the
$\alpha^{th}$ point defect, which has a topological charge $q_{\alpha}$.
We will restrict the analysis here to the case of charge $\pm 1$
vortices where $q_{\alpha}^{2}=1$.  This case dominates the late-stage
ordering since higher-charged defects are energetically unfavourable
and dissociate into charge $\pm 1$  defects early on.

The next step is to note, as pointed out by Halperin \cite{HALPERIN81},
that the positions of defects are located by the zeros of the order-parameter
field $\vec{ \psi}$. Therefore the charged or signed density
for point defects is given by
%
% DENSITY IN TERMS OF THE ORDER PARAMETER
%
\be
\rho (1)=\delta[\vec{ \psi}(1)]{\cal D}(1)
\label{EQ:SIGNEDDENSITY}
\ee
where the Jacobian ${\cal D}$ associated with the change of variables
from the set of vortex positions to the field $\vec{\psi}$ is defined by:
%
% DETERMINANT
%
\be
{\cal D}(1) =\frac{1}{n!}\epsilon_{\mu_{1},\mu_{2},...,\mu_{n}}
\epsilon_{\nu_{1},\nu_{2},...,\nu_{n}}
\nabla_{\mu_{1}}\psi_{\nu_{1}}
\nabla_{\mu_{2}}\psi_{\nu_{2}}...
\nabla_{\mu_{n}}\psi_{\nu_{n}}
\label{EQ:DET}
\ee
where $\epsilon_{\mu_{1},\mu_{2},...,\mu_{n}}$ is the $n$-dimensional fully
anti-symmetric tensor and summation over repeated indices is implied.
The unsigned  density, $n(1)$, does not consider the charge of the defect
and is given by
%
% UNSIGNED DEFECT DENSITY
%
\bea
n (1) & = & \sum_{\alpha} \delta( {\bf r}_{1} - {\bf x}_{\alpha} (t_{1}) )
\nonumber \\
& = & \delta[\vec{ \psi}(1)]|{\cal D}(1)|.
\label{EQ:UNDEF}
\eea
If we have products of such densities at equal-times,
$t_{1}=t_{2}=t$, we write
%
% PRODUCT FORM
%
\be
\label{EQ:PRODFORMO}
\rho (1) \rho (2)=\delta ({\bf r}_{1}-{\bf r}_{2}) \sum_{\alpha}
\delta({\bf r_{1}} - {\bf x_{\alpha}}(t))
+\tilde{\rho} (1) \tilde{\rho} (2).
\ee
We use the tildes in (\ref{EQ:PRODFORMO}) to indicate that the product
$\tilde{\rho} (1)\tilde{\rho} (2)$ contains only terms arising from different
defects. One may write
%
% PRODUCT FORM: VERSION 2
%
\be
\label{EQ:PRODFORM}
\rho (1)\rho (2)=\delta ({\bf r}_{1}-{\bf r}_{2}) n(1)
+\tilde{\rho} (1) \tilde{\rho} (2).
\ee
The equal-time
charged or signed defect correlation function is given by the average of
(\ref{EQ:PRODFORM})
%
% SIGNED VORTEX CORRELATIONS
%
\bea
G_{s}({\bf r},t) & =  & \langle \rho(1)\rho(2) \rangle \nonumber \\
                 & =  & n_{0} (t)
\mbox{ } \delta({\bf r})+ \tilde{G}_{s}({\bf r},t).
\label{EQ:SIGNVORCOR}
\eea
The first term in (\ref{EQ:SIGNVORCOR}) represents self-correlations and
%
% DEFINITION OF THE DEFECT DENSITY
%
\be
n_{0} (t)=\langle n(1) \rangle
\ee
is the the average unsigned point defect density at time $t$ \cite{NEUTRALITY}.
The second term in (\ref{EQ:SIGNVORCOR}) is
%
% SIGNED VORTEX CORRELATIONS
%
\be
\label{EQ:TSVORCOR}
\tilde{G}_{s}({\bf r},t)= \langle \tilde{\rho}(1) \tilde{\rho}(2) \rangle
\ee
and measures the signed correlation between different defects.
It is easy to see that the signed defect correlation function, $G_{s}$,
can also be decomposed as
%
% SECOND DECOMPOSITION
%
\be
\label{EQ:GSVV}
G_{s} = 2 C_{vv}-2 C_{va}
\ee
where $C_{vv}$ is the correlation function between like-signed
vortices and $C_{va}$ is the correlation function between vortices and
antivortices.

We can also define the equal-time unsigned defect correlation
function as
%
% UNSIGNED VORTEX CORRELATIONS
%
\bea
G_{u}({\bf r},t) & = & \langle  n(1)n(2) \rangle \nonumber \\
                 & = & n_{0} (t) \mbox{ }
\delta({\bf r})+ \tilde{G}_{u}({\bf r},t)
\label{EQ:UNSIGNVORCOR}
\eea
with
%
% DEFINITION OF TILDE GU
%
\be
\label{EQ:TILDEGU}
\tilde{G}_{u}({\bf r},t)= \langle \tilde{n}(1)\tilde{n}(2) \rangle.
\ee
As with the signed quantity, one can write $G_{u}$ in terms of the
vortex-vortex and vortex-antivortex correlation functions:
%
% GU IN TERMS OF CVV AND CVA
%
\be
\label{EQ:GUVV}
G_{u}= 2 C_{vv}+2 C_{va}.
\ee
Inverting (\ref{EQ:GSVV}) and (\ref{EQ:GUVV}) and using
(\ref{EQ:SIGNVORCOR}) and (\ref{EQ:UNSIGNVORCOR}) one has
%
% EXPLICIT FORMS FOR CVV AND CVA
%
\bea
C_{vv} ({\bf r},t) & = &
\frac{1}{2}  n_{0}(t) \delta ({\bf r}) + \tilde{C}_{vv} ({\bf r},t) \\
C_{va} ({\bf r},t) & = & \frac{1}{4} [
\tilde{G}_{u}({\bf r},t) - \tilde{G}_{s}({\bf r},t)],
\eea
where
%
% TILDE CVV DEFN
%
\be
\tilde{C}_{vv} ({\bf r},t) = \frac{1}{4} [
\tilde{G}_{s} ({\bf r},t) + \tilde{G}_{u} ({\bf r},t)].
\ee
As one would expect, for the vortex-antivortex correlations, there is no
$\delta$-function contribution from self-correlations.

We are interested in the correlations between different vortices and
antivortices in the scaling regime and we can define the following scaling
forms for the quantities of interest:
%
% SIGNED AND UNSIGNED DEFECT SCALING FUNCTIONS - VV AND VA
%
\bea
\label{EQ:GSSCAL}
{\cal G}_{s} (x) & \equiv & \frac{\tilde{G}_{s}({\bf r},t)}{[n_{0}(t)]^{2}} \\
{\cal G}_{u} (x) & \equiv & \frac{\tilde{G}_{u}({\bf r},t)}{[n_{0}(t)]^{2}}
\label{EQ:GUSCAL} \\
{\cal C}_{vv} (x) & \equiv  & \frac{4 \tilde{C}_{vv} ({\bf r},t)}
{[n_{0}(t)]^{2}} =
{\cal G}_{s} (x) + {\cal G}_{u} (x)  \label{EQ:CVVSCAL} \\
{\cal C}_{va} (x) & \equiv & \frac{4 C_{va} ({\bf r},t)}{ [n_{0}(t)]^{2} }
= {\cal G}_{u} (x) - {\cal G}_{s} (x). \label{EQ:CVASCAL}
\eea
Both ${\cal C}_{vv} (x)$ and ${\cal C}_{va} (x)$ are normalized to
approach $1$ as $x \rightarrow \infty$.

Since the topological charge is conserved, one has the very important
constraint \cite{HALPERIN81,NEUTRALITY}
%
% CONSERVATION LAW
%
\be
\int ~d^{d}r \mbox{ } G_{s}({\bf r},t)=0.
\ee
Using (\ref{EQ:SIGNVORCOR}) this conservation law can be written in the form
%
% CONSERVATION LAW: VERSION 2
%
\be
\int ~d^{d}r~\tilde{G}_{s} ({\bf r},t)= - n_{0}(t).
\label{EQ:CONSLAW}
\ee
Theory, simulation, and experiment indicate that the scaling
results (\ref{EQ:GSSCAL}) - (\ref{EQ:CVASCAL}) and
%
% DENSITY SCALING RELATION
%
\be
\label{EQ:DENSCALREL}
n_{0}(t)=\frac{A}{L^{d}(t)},
\ee
where $A$ is a constant, hold. Inserting these results in (\ref{EQ:CONSLAW})
leads to the relation
%
% RELATION BETWEEN A AND GS
%
\be
\label{EQ:SUMRULE}
\int ~d^{d}x~ {\cal G}_{s}(x)=-\frac{1}{A}.
\ee
A measurement of $n_{0}(t)$ and a choice for $L(t)$ fixes A, allowing one to
check that the sum rule (\ref{EQ:SUMRULE}) is satisfied. As shown
in \cite{LIU92b} our theoretical results satisfy this sum rule exactly.

The results presented above are rather general. To
evaluate $G_{s}$ explicitly one can use the gaussian closure approximation, as
was done in \cite{LIU92b}.
The evaluation of $G_{u}$ in this same approximation
is technically much more difficult than the calculation of $G_{s}$
because of the absolute value sign in the definition (\ref{EQ:UNDEF}) of the
unsigned defect density $n(1)$. The purpose of this paper is to compute
$G_{u}$ and thereby obtain $C_{vv}$ and $C_{va}$.
In sections VI and VII we carry out this calculation with explicit
results for $n=d=2$. First, however, we must briefly review the calculations
of the order-parameter scaling function and the signed defect
correlation function within the gaussian closure approximation.
%
% THE GAUSSIAN CLOSURE APPROXIMATION
%
\section{The Gaussian Closure Approximation}

Substantial progress has been made in determining the
order-parameter scaling function ${\cal F}$ using the theory
developed in \cite{MAZENKO90}.
In this and related theories, one
expresses the order parameter $\OP({\bf r},t)$ as a local non-linear function
of an auxiliary field
$\vec{m}({\bf r},t)$ which is physically interpreted as the distance, at time
$t$, from position ${\bf r}$ to the closest defect.
One of the physical motivations for introducing $\vec{m}({\bf r},t)$
is that it is {\it smoother} than the order-parameter field.  Sharp
interfaces or well-defined defects produce a non-analytic structure in
the order-parameter scaling function $\F(x)$
at small-scaled distances $x$ which is responsible for
the Porod's law decay seen scattering experiments \cite{POROD}. The
expectation, however, is that the auxiliary field correlation
function $f(x)$, defined below,
 will be analytic in this same distance range.  In the case of
a scalar order-parameter these expectations are supported by the
theory \cite{MAZENKO90}. However, for the simplest theory \cite{LIU92a}
with $n > 1$ this is not the case.
One finds a weak  non-analytic component in
$f$ and, more significantly, for $n=2$ one can trace this non-analytic
component to an unphysical divergence \cite{LIU92b} in the scaling form for
the signed vortex correlation function ${\cal G}_{s}(x)$ at small $x$.
In previous work \cite{MAZENKO96} for  $n=2$  we showed
how these problems can be resolved by taking seriously the
assumption that the correlations of the auxiliary field
are indeed smoother than those of the order parameter.
We find that it is possible to rearrange the theory such that
$f$ is analytic in $x$ if we extend the theory to include fluctuations
about the ordering field and treat the separation between the ordering
field and the fluctuation field carefully.

More specifically we can decompose the order parameter $\OP$  as
%
% MAPPING
\be
\label{EQ:MAPPING}
\OP = \SG [ \M ] + \U.
\ee
$\SG$ is chosen to reflect the defect
structure in the problem and is of ${\cal O}(1)$.
$\U$ represents fluctuations about the ordering field $\SG$
and is of ${\cal O}(L^{-2})$ at late times.
The defect structure is incorporated
by demanding that $\SG$ satisfy the
Euler-Lagrange equation for the order-parameter around a static defect in
equilibrium. This determines $\SG$ as a function of $\M$. Since we
expect only the lowest-energy defects, having unit topological charge,
will survive to late-times we obtain \cite{LIU92a}
%
% EXPLICIT FORM FOR SIGMA
\be
\SG [\M] = A(m) \hat{m}.
\label{EQ:SIGEX}
\ee
In \cite{LIU92a} it was shown that $A$ increases
linearly from zero near the defect core and relaxes algebraically to its
ordered value $A=\psi_{0}$ for large $m \equiv |\M|$.

Evolution under (\ref{EQ:MOT}) causes $\OP$ to order and assume a
distribution that is far from gaussian. However, it is reasonable to assume
that the
the probability distribution for the auxiliary field $\M$ will be near a
gaussian. Indeed, a simple and successful assumption \cite{POSTGAUSSIAN}
to make is that the probability distribution for $\M$ is gaussian with
the correlation function $C_{0} (12)$  explicitly defined through
%
% DEFINITION OF C0
\be
\langle m_{i} (1) m_{j} (2) \rangle = \delta_{ij} \mbox{ } C_{0} (12).
\ee
The system is assumed to be statistically isotropic and homogeneous so
$C_{0}(12)$ is invariant under interchange of its spatial indices. For future
reference we also define the one-point correlation function
%
% DEFINITION OF S0
%
\be
S_{0}(1) = C_{0}(11)
\ee
and the normalized correlation function
%
% DEFINITION OF f
%
\be
f(12) = \frac{C_{0}(12)}{S_{0}(1)}.
\ee
Since $m$ measures the characteristic distance between defects it is
expected to grow as $L$ in the late-time scaling regime.
This means that  $C_{0}$ and $S_{0}$ grow as $L^{2}$ at late times.

If $\vec{m}$ is treated as a gaussian variable then its probability
distribution is characterized by the single function $f$.
This function can be determined by requiring that the
equation of motion for $\SG$ be satisfied on average.
In \cite{MAZENKO96} it was shown that for $n=2$ this requirement produces the
following late-time scaling equation for the equal-time order-parameter
correlations:
%
% SCALING EQUATION
%
\be
\label{EQ:SCALE}
\vec{x} \cdot \nabla_{x} \F + \nabla_{x}^{2} \F - \frac{\pi}{4 \mu} \F +
\frac{\pi}{2 \mu}
f \partial_{f} \F
= 0
\ee
where, for general $n$, ${\cal F}$ (\ref{EQ:OPSCAL}) is related to
$f$ {\em via}
%
% HYPERGEOMETRIC RELATION
%
\be
\label{EQ:HYPER}
\F = \frac{n}{2 \pi} B^{2} \left[ \frac{1}{2},\frac{n+1}{2} \right]
F \left[ \frac{1}{2}, \frac{1}{2};\frac{n+2}{2};f^{2} \right].
\ee
Here $B$ is the beta function and $F$ is the hypergeometric function
\cite{LIU92a,BRAY91}.
In the derivation of (\ref{EQ:SCALE}) we have  defined the scaling length
%
% LENGTH DEFINITION
%
\be
\label{EQ:LENGTH}
L^{2}(t) = \frac{\pi S_{0}(t)}{2 \mu} = 4 t.
\ee
With the theory in this form there will be no leading small-$x$
non-analyticities in the normalized auxiliary field correlation function $f$
and the small-$x$ divergence in the scaling form for the
signed vortex correlation function found in earlier theories does not appear.
The calculation of the scaling form for $\F$ (and thus $f$)
reduces to the solution of
the non-linear eigenvalue problem (\ref{EQ:SCALE}) with the eigenvalue $\mu$.
The eigenvalue is selected by numerically  \cite{MAZENKO96}
finding the solution of
(\ref{EQ:SCALE}) which satisfies the analytically determined boundary behaviour
at both large and small $x$.
%
% THE SIGNED VORTEX CORRELATION FUNCTION
%
\section{THE SIGNED DEFECT CORRELATION FUNCTION}

The calculation of $G_{s}({\bf r},t)$ (\ref{EQ:SIGNVORCOR}),
carried out in \cite{LIU92b}, begins
with the observation that $\vec{\psi}$ and $\M$ share the same
zeros, and that near these zeros we can use (\ref{EQ:SIGEX}) to write
%
% SMALL M FORM FOR PSI
%
\be
\vec{\psi}=a_{0}\vec{m}+b_{0}m^{2}\vec{m}+...
\ee
where $a_{0}$ and $b_{0}$ are constants which depend on the potential.
It is then easy to see that in equations (\ref{EQ:SIGNEDDENSITY}) and
(\ref{EQ:UNDEF}) for  $\rho (1)$ and $n (1)$ we can replace $\vec{ \psi}(1)$
with $\M(1)$ and the factors of $a_{0}$ and $b_{0}$ all cancel.
Then, assuming $\M$ is a gaussian field,  it is straightforward to
see that $\tilde{G}_{s}({\bf r},t)$ (\ref{EQ:TSVORCOR}) factors into a product
of gaussian averages which can be evaluated using standard methods.
One then finds that $\tilde{G}_{s}({\bf r},t)$ indeed has the scaling form
(\ref{EQ:GSSCAL}) with ${\cal G}_{s} (x)$ given by \cite{NOTATION}
%
% DEFECT DEFECT CORRELATION FUNCTION
%
\be
\label{EQ:DEFECTDEFECTCOR}
{\cal G}_{s} (x) = \frac{\Gamma^{2} (1 + n/2)}{n!} \left( \frac{8 \mu}{S^{(2)}}
\right)^{n} \left[ \frac{h(x)}{x} \right]^{n-1}
\frac{\partial h(x)}{\partial x}
\ee
with
%
% H DEFN
%
\bea
\label{EQ:HDEF}
h  & = & - \frac{\gamma f'}{2 \pi} \\
\gamma & = & \frac{1}{\sqrt{1-f^2}},
\eea
and
%
% S2 DEFN
%
\be
S^{(2)} \equiv \frac{1}{n^{2}} \langle [ \vec{\nabla} \M ]^{2} \rangle =
\frac{1}{n}
\ee
in this theory. The defect density is given by
%
% DEFECT DENSITY
%
\be
\label{EQ:DEFDENS}
n_{0}(t) = \frac{n!}{2^{n/2}\Gamma(1 + n/2)} \left[
\frac{S^{(2)}}{2\pi S_{0}(t)} \right]^{n/2}.
\ee
Equation (\ref{EQ:DEFDENS}) leads to the
result $n_0 \sim
L^{-n}$ which is just a restatement
that there is scaling in the problem governed by the single
length $L(t)$.

Since $f$ is determined in the theory
for the order-parameter correlation function, the function
$G_{s}({\bf r},t)$ is fully
determined in the scaling regime. The derivatives
in (\ref{EQ:DEFECTDEFECTCOR}) and (\ref{EQ:HDEF}) make ${\cal G}_{s}(x)$
 a rather sensitive function of $f(x)$ for small $x$.

\section{Calculation of $\tilde{G}_{u}$}

The equal-time unsigned defect correlation function $G_{u}({\bf r},t)$
(\ref{EQ:UNSIGNVORCOR}) can be evaluated through a series of steps.
We work with general $n$. The average that needs to be
computed is (\ref{EQ:TILDEGU})
%
% TILDE GU AVERAGE
%
\be
\label{EQ:TGUAVG}
\tilde{G}_{u}({\bf r},t) = \langle \delta[\psi (1) ] | {\cal D} (1)|
\delta[\psi (2)] | {\cal D} (2) | \rangle.
\ee
The first step is to realize that
one can replace $\vec{\psi}$ by $\vec{m}$ in (\ref{EQ:TGUAVG}) and write
$\tilde{G}_{u}$
in terms of integrals over  the reduced
probability distribution $G(\xi_{1},\xi_{2})$:
%
% INTEGRAL FORM FOR GU
%
\be
\label{EQ:GUINT}
\tilde{G}_{u}({\bf r},t)=\int ~\prod_{\mu \nu}d(\xi_{1})_{\mu}^{\nu}
d (\xi_{2})_{\mu}^{\nu} |{\cal D}(\xi_{1})||{\cal D}(\xi_{2})|
G(\xi_{1},\xi_{2}),
\ee
where
%
% DETERMINANT
%
\be
{\cal D}(\xi) =\frac{1}{n!}\epsilon_{\mu_{1},\mu_{2},...,\mu_{n}}
\epsilon_{\nu_{1},\nu_{2},...,\nu_{n}}
\xi_{\mu_{1}}^{\nu_{1}}
\xi_{\mu_{2}}^{\nu_{2}}....
\xi_{\mu_{n}}^{\nu_{n}}
\ee
and
%
% G XI
%
\be
G(\xi_{1},\xi_{2}) = \langle \delta[\M(1)] \delta[\M(2)] \prod_{\mu \nu}
         \delta[(\xi_{1})_{\mu}^{\nu} - \nabla_{\mu} m_{\nu} (1) ]
         \delta[(\xi_{2})_{\mu}^{\nu} - \nabla_{\mu} m_{\nu} (2) ]
\rangle.
\ee
The indices $\mu$ and $\nu$ range from $1$ to $2$, unless stated otherwise.
The gaussian average defining $G(\xi_{1},\xi_{2})$
is calculated in Appendix A.
It is shown there how one can write $G(\xi_{1},\xi_{2})$
in terms of the longitudinal and
transverse components of the rotated variables  $(t_{j})_{\mu}^{\nu}$ $(j = 1
\mbox{ or } 2)$ defined by
%
% ROTATION
%
\be
(t_{j})_{\mu}^{\nu} = \hat{M}_{\beta}^{\mu} \mbox{ } (\xi_{j})_{\beta}^{\nu}
\ee
where $\hat{M}$ is an orthogonal matrix. We define the longitudinal piece of
$t_{j}$ as
$(t_{j})_{L}^{\nu} \equiv (t_{j})_{1}^{\nu} $ and write $(t_{j})_{T}$ to
denote the transverse pieces: $(t_{j})_{\mu}^{\nu}$ with $\mu > 1$.
One then obtains:
%
% G(t)
%
\be
G(t_{1},t_{2}) =
\left( \frac{\gamma}{2 \pi S_{0}} \right)^{n} G_{L}((\vec{t}_{1})_{L},
(\vec{t}_{2})_{L}) \mbox{ } G_{T}((t_{1})_{T},(t_{2})_{T})
\ee
where
%
% LONGITUDINAL AND TRANSVERSE G
%
\bea
\label{EQ:D1}
G_{L} ((\vec{t}_{1})_{L},(\vec{t}_{2})_{L}) & = &
\left( \frac{\gamma_{L}}{2 \pi S_{L}} \right)^{n}
\exp - \frac{\gamma_{L}^{2} }{2 S_{L}} [ \mbox{ } \sum_{j}
(\vec{t}_{j})_{L}^{2} - 2 f_{L} \mbox{ } (\vec{t}_{1})_{L} \cdot
(\vec{t}_{2})_{L} \mbox{ } ] \\
G_{T} ((t_{1})_{T},(t_{2})_{T}) & = &
\left( \frac{\gamma_{T}}{2 \pi S_{T}} \right)^{n(n-1)}
\exp - \frac{\gamma_{T}^{2} }{2 S_{T}} [  \mbox{ }
\sum_{\mu = 2} \sum_{\nu j}
[(t_{j})_{\mu}^{\nu}] ^{2} - 2 f_{T} \sum_{\mu = 2} \sum_{\nu j}
(t_{1})_{\mu}^{\nu} (t_{2})_{\mu}^{\nu} \mbox{ } ].
\label{EQ:D2}
\eea
The longitudinal
quantities $S_{L}$, $f_{L}$ and $\gamma_{L}$ are defined in terms of
$C_{0}$ and $S_{0}$ through
%
% LONGITUDINAL QUANTITIES
%
\bea
S_{L} & = & S^{(2)} - \frac{\gamma^{2}}{S_{0}} ( C_{0}' )^{2}
\label{EQ:SL} \\
C_{L} & = & - C_{0} '' - \frac{f \gamma^{2} }{S_{0}}  ( C_{0}' )^{2}
\label{EQ:CL} \\
f_{L} & = & \frac{C_{L}}{S_{L}} \label{EQ:FL} \\
\gamma_{L}^{2} & = & (1 - f_{L}^{2} ) ^{-1}.  \label{EQ:GAMMAL}
\eea
The corresponding transverse functions are
%
% TRANSVERSE QUANTITIES
%
\bea
S_{T} & = & S^{(2)} \label{EQ:ST} \\
C_{T} & = & - \frac{C_{0}'}{r} \label{EQ:CT} \\
f_{T} & = & \frac{C_{T}}{S_{T}}  \label{EQ:FT} \\
\gamma_{T}^{2} & = & (1 - f_{T}^{2}) ^{-1}. \label{EQ:GAMMAT}
\eea
Since the matrix $\hat{M}$ is orthogonal we have
%
% DETERMINANT RELATIONS
%
\be
{\cal D}(\xi_{j}) ={\cal D}(t_{j})
\ee
so we may write
%
% DECOMPOSED FORM FOR GU
%
\be
\tilde{G}_{u}({\bf r},t)=\int ~\prod_{\mu\nu} d (t_{1})_{\mu}^{\nu}
d (t_{2})_{\mu}^{\nu}
|{\cal D}(t_{1})||{\cal D}(t_{2})|
\left( \frac{\gamma}{2 \pi S_{0}} \right)^{n}
G_{L} ((\vec{t}_{1})_{L},(\vec{t}_{2})_{L}) \mbox{ } G_{T} ( (t_{1})_{T},
 (t_{2})_{T}).
\ee
Under the change of variables
%
% CHANGE OF VARIABLES
%
\be
(\vec{t}_{j})_{L} =\sqrt{\frac{S_{L}}{\gamma_{L}^{2}}} \mbox{ }
(\vec{s}_{j})_{L}
\ee
and, for $\mu > 1$,
\be
(t_{j})_{\mu}^{\nu}=\sqrt{\frac{S_{T}}{\gamma_{T}^{2}}} \mbox{ }
(s_{j})_{\mu}^{\nu}
\ee
$\tilde{G}_{u}({\bf r},t)$ becomes
%
% FINAL FORM FOR GU
%
\be
\label{EQ:GUFINAL}
\tilde{G}_{u}({\bf r},t)=\left( \frac{\gamma}{2 \pi S_{0}} \right)^{n}
\frac{S_{L}(S_{T})^{n-1}}{(2\pi )^{n^{2}}}
\gamma_{L}^{-(n+2)}\gamma_{T}^{-(n+2)(n-1)} \mbox{ } {\cal N}(f_{T},f_{L})
\ee
where
%
% CAL N DEFN
%
\bea
\lefteqn{
{\cal N}(f_{T},f_{L})=\int ~\prod_{\mu \nu}d (s_{1})_{\mu}^{\nu}
d (s_{2})_{\mu}^{\nu}
|{\cal D}(s_{1})||{\cal D}(s_{2})|
\exp -\frac{1}{2} [ (\vec{s}_{1})_{L}^{2} +
(\vec{s}_{2})_{L}^{2}
-2f_{L}\mbox{ } (\vec{s}_{1})_{L} \cdot (\vec{s}_{2})_{L}  ]
} \nonumber \\ & &
\hspace{1.5 in} \times \exp -\frac{1}{2} \sum_{\mu = 2} \sum_{\nu}
( [(s_{1})_{\mu}^{\nu}]^{2}
+[(s_{2})_{\mu}^{\nu}]^{2}
-2f_{T} \mbox{ } (s_{1})_{\mu}^{\nu} (s_{2})_{\mu}^{\nu} ). \label{EQ:CALN}
\eea
Equation
(\ref{EQ:GUFINAL}), with the definition (\ref{EQ:CALN}),
 is one of the central results of this
paper. If
one took this route in evaluating $\tilde{G}_{s}$ one would arrive at these
same equations, only without the absolute value signs in (\ref{EQ:CALN}).
In the next
section we will examine the $n=d=2$ case and go further to derive an
expression for $\tilde{G}_{u}$ that is convenient for numerical work.

However, first we examine the general expression (\ref{EQ:GUFINAL}) for
$\tilde{G}_{u}$ in the small-$x$ limit.
We show in Appendix B that, in the scaling regime, as $x\rightarrow 0$,
$f_{T} \rightarrow 1$ and, surprisingly,
$f_{L} \rightarrow -1$.  To examine this limit for $\tilde{G}_{u}$ we make
the change of variables
%
% TRANSVERSE CHANGE
%
\be
(s_{1})_{\mu}^{\nu} =\frac{1}{\sqrt{\epsilon_{T}}}\phi_{\mu}^{\nu}
+\frac{1}{2}\chi_{\mu}^{\nu}
\ee
\be
(s_{2})_{\mu}^{\nu} =\frac{1}{\sqrt{\epsilon_{T}}}\phi_{\mu}^{\nu}
-\frac{1}{2}\chi_{\mu}^{\nu}
\ee
for $\mu > 1$ with
\be
\epsilon_{T}=2(1-f_{T})
\ee
and, for $\mu=1$,
\be
(s_{1})_{L}^{\nu} =-\frac{1}{\sqrt{\epsilon_{L}}}\phi_{L}^{\nu}
+\frac{1}{2}\chi_{L}^{\nu}
\ee
\be
(s_{2})_{L}^{\nu} =\frac{1}{\sqrt{\epsilon_{L}}}\phi_{L}^{\nu}
+\frac{1}{2}\chi_{L}^{\nu}
\ee
with
\be
\epsilon_{L}=2(1+f_{L}).
\ee
Then as $\epsilon_{T}$ and $\epsilon_{L}$ go to zero the
arguments of the exponentials in (\ref{EQ:CALN}) go over to
%
% EXPONENTIAL ARGUEMENTS
%
\be
[(s_{1})_{\mu}^{\nu}]^{2}+[(s_{2})_{\mu}^{\nu}]^{2}
-2f_{T} \mbox{ } (s_{1})_{\mu}^{\nu} (s_{2})_{\mu}^{\nu}
=[\phi_{\mu}^{\nu}]^{2}+[\chi_{\mu}^{\nu}]^{2}
\ee
for $\mu > 1$ and
\be
(\vec{s}_{1})_{L}^{2} + (\vec{s}_{2})_{L}^{2}
-2f_{L} \mbox{ } (\vec{s}_{1})_{L} \cdot (\vec{s}_{2})_{L}
=[\vec{\phi}_{L}]^{2}+[\vec{\chi}_{L}]^{2}
\ee
for $\mu = 1$.
The important point is that, as $\epsilon_{L}$ and $\epsilon_{T}$ go to zero,
the Jacobians ${\cal D}$ transform as
\bea
{\cal D}(s_{1}) & = & \frac{-1}{\sqrt{\epsilon_{L}}}
\frac{1}{(\epsilon_{T})^{\frac{n-1}{2}}}{\cal D}(\phi) \\
{\cal D}(s_{2}) & = & \frac{1}{\sqrt{\epsilon_{L}}}
\frac{1}{(\epsilon_{T})^{\frac{n-1}{2}}}{\cal D}(\phi).
\eea
Thus the Jacobians differ only by a sign in this limit and we have
%
% SHORT DISTANCE JACOBIAN RELATIONS
%
\be
|{\cal D}(s_{1})||{\cal D}(s_{2})|=-{\cal D}(s_{1}){\cal D}(s_{2})
\ee
as $x \rightarrow 0$.
Since the scaling form for the
unsigned defect correlation function ${\cal G}_{s}(x)$ is, up to a factor
of $[n_{0}(t)]^{2}$, given by
(\ref{EQ:GUFINAL}) and (\ref{EQ:CALN}) without the absolute value signs
we see that ${\cal G}_{u}(x)$ differs from ${\cal G}_{s}(x)$ only by
a sign
%
% SHORT DISTANCE RELATION BETWEEN GU AND GS
%
\be
{\cal G}_{u}=-{\cal G}_{s}
\ee
as $x\rightarrow 0$.  The relations (\ref{EQ:CVVSCAL}) and
(\ref{EQ:CVASCAL}) then lead to the results
%
% SHORT CVV
%
\be
\lim_{x\rightarrow 0} {\cal C}_{vv}(x) = 0
\ee
and
%
% SHORT CVA
%
\be
\lim_{x\rightarrow 0} {\cal C}_{va}(x)= -2 {\cal G}_{s} (0).
\ee
Thus there is a depletion zone at short-scaled distances
for like-signed defects.  From previous work we know that
${\cal G}_{s} (0) < 0$ \cite{MAZENKO96} so the theory
gives a non-zero, positive correlation at short-scaled distances
for unlike-signed defects. These are general results and
depend only on the gaussian assumption, used to derive (\ref{EQ:GUFINAL}),
and the particular small-$x$ behaviour of $f$ which, as is shown in
Appendix C, determines the small-$x$ behaviour of
the quantities $f_{T}$ and $f_{L}$.
%
% TWO-DIMENSIONAL O(2) MODEL
%
\section{TWO-DIMENSIONAL $O(2)$ MODEL}

We have not yet been able to explicitly
evaluate ${\cal G}_{u}(x)$ for general $n=d$, except for small and large
$x$.  Here we
specialize to $n=d=2$.  Much theoretical work has recently been done on this
case and detailed results are available for the auxiliary field correlation
function $f(x)$. An
additional motivation for examining this case is that simulation and
experimental results exist for the vortex-vortex and vortex-antivortex
correlation functions.

If we now specialize (\ref{EQ:GUFINAL}) and (\ref{EQ:CALN})
to $n=d=2$ then the unsigned vortex correlation function is
given by
%
% n=d=2 UNSIGNED CORRELATION FUNCTION
%
\be
G_{u}({\bf r},t) = [n_{0}(t)]^{2}
\frac{2 \gamma^{2} S_{L}}{ (2\pi )^{4}}
\gamma_{L}^{-4}\gamma_{T}^{-4}{\cal N}(f_{T},f_{L})
\ee
with
%
% DEFINITION OF N
%
\be
\label{EQ:N}
{\cal N}(f_{T},f_{L})=\int ~d^{2} s_{1} d^{2} s_{2} \mbox{ }
J(\vec{s}_{1},\vec{s}_{2}) \mbox{ }
\exp -\frac{1}{2} [ {\vec{s}_{1}}^{2} + {\vec{s}_{2}}^{2}
-2 f_{L} \mbox{ } \vec{s}_{1} \cdot \vec{s}_{2} ],
\ee
where
%
% DEFINITION OF J
%
\be
J(\vec{s}_{1},\vec{s}_{2}) = \int d^{2} x d^{2} y
| s_{11} x_{2} - s_{12} x_{1} || s_{21} y_{2} - s_{22} y_{1} |
\exp - \frac{1}{2} [ \vec{x} \mbox{}^{2} + \vec{y} \mbox{}^{2}
 - 2 f_{T} \mbox{ }
\vec{x} \cdot \vec{y} \mbox{ }  ].
\label{EQ:DEFNJ}
\ee
We have simplified the notation by writing $\vec{s}_{i}$ in place of
$(\vec{s}_{i})_{L}$ and using $\vec{x}$ and $\vec{y}$ in place of
$(\vec{s}_{1})_{T}$ and $(\vec{s}_{2})_{T}$ respectively.
The $j^{th}$ component of $\vec{s}_{i}$ is written $s_{ij}$.
The quantity $J$ is evaluated in Appendix B with the clean result:
%
% FINAL FORM FOR J
%
\be
J(\vec{s}_{1},\vec{s}_{2})= 2 \pi \gamma_{T}^{4}|\vec{s}_{1}||\vec{s}_{2}|
\tilde{J}(g)
\ee
where
%
% DEFINITION OF TILDE J
%
\be
\tilde{J}(g) = 4 \sqrt{1-g} + 4 \sqrt{g} \tan^{-1} \sqrt{\frac{g}{1-g}}
\ee
and
%
% DEFINITION OF G
%
\be
g=f_{T}^{2} (\hat{s}_{1} \cdot \hat{s}_{2})^{2}.
\ee
Notice that $g$ depends only on the angle between $\vec{s}_{1}$
and $\vec{s}_{2}$ and not on their magnitudes.  We can then separate
(\ref{EQ:N}) into an integration over magnitudes, followed by an overall
integration over the angular piece:
%
% SEPARATION OF INTEGRATIONS
%
\be
{\cal N}(f_{T},f_{L}) = (2 \pi)^{2} \gamma_{T}^{4}
 \int_{0}^{2 \pi} d \theta \tilde{J}(g) \int_{0}^{\infty} d s_{1} d s_{2}
\mbox{ } s_{1}^{2} s_{2}^{2} \exp -\frac{1}{2}[ s_{1}^{2} + s_{2}^{2}
- 2 g_{L} \mbox{ } s_{1} s_{2} ]
\label{EQ:NSEP}
\ee
where we have defined
%
% DEFINITION OF GL
%
\be
g_{L} = f_{L} \mbox{ } \hat{s}_{1} \cdot \hat{s}_{2}
\ee
and $\hat{s}_{1} \cdot \hat{s}_{2} = \cos \theta$. Expanding the exponential
in (\ref{EQ:NSEP}) as a power series in $g_{L}$ and performing the
integrations over the magnitudes $s_{1}$ and $s_{2}$ we obtain
%
% FINAL FORM FOR N
%
\be
{\cal N}(f_{T},f_{L}) = (2 \pi)^{2} \gamma_{T}^{4}
\int_{0}^{2 \pi} d \theta \tilde{J}(g) \sum_{\ell = 0}^{\infty}
\frac{ (g_{L})^{\ell} }{ \ell !} \mbox{ } 2^{\ell + 1} \mbox{ }
\Gamma^{2} \left( \frac{\ell + 3}{2} \right).
\label{EQ:FINN}
\ee
The symmetries of
$g$ and $g_{L}$ under $\theta \rightarrow \theta - \pi$ allow us to restrict
the region of integration in (\ref{EQ:FINN}) and we use the definition
(\ref{EQ:GUSCAL}) for the scaling form to write the final result for
${\cal G}_{u}$:
\be
{\cal G}_{u}(x) = \frac{2 \gamma^{2} S_{L} \gamma_{L}^{-4}}{\pi^{2}}
\int_{0}^{\pi/2} d \theta \tilde{J}(g) \sum_{\ell = 0}^{\infty} \frac{(g_{L})
^{2 \ell} }{(2 \ell) !} \mbox{ }
2^{2 \ell + 1} \mbox{ } \Gamma^{2} \left( \frac{2 \ell + 3}{2}
\right)
\label{EQ:GUNUM}
\ee
where $S_{L}$, $f_{L}$ and $f_{T}$
are now functions of the scaling variable $x$.
The integral over $\theta$ and sum over $\ell$ have to be evaluated
numerically.

For large $x$ the functions $f$, $f_{T}$ and $f_{L}$ are all small and and
(\ref{EQ:GUNUM}) simplifies since only
the first two terms in the series need to be retained to give the essential
physical features. In this limit the integral over $\theta$ is easily
performed and one has
%
% LARGE X - GU
%
\be
{\cal G}_{u} = 1 + f^{2} + \frac{1}{4} ( f_{L}^{2} + f_{T}^{2} )
- \frac{4 \mu}{\pi} (f')^{2}.
\ee
Since all of $f$, $f_{T}$ and $f_{L}$ decay as $e^{-x^{2}/2}$ for large $x$
\cite{MAZENKO96}, ${\cal G}_{u}$ rapidly approaches $1$ as $x$ increases.
%
% COMPARISON OF SIMULATION AND THEORY
%
\section{Comparison of Theory and Simulation}

In our previous work \cite{MAZENKO96} we numerically solved the eigenvalue
problem (\ref{EQ:SCALE}) and determined the function $f(x)$
representing the scaling form for correlations in the
auxiliary field. With this information we can use (\ref{EQ:GUNUM}) to
determine ${\cal G}_{u}(x)$ since for each $x$ at which we know $f(x)$
we can calculate $f_{L}$, $f_{T}$ and $S_{L}$ and perform the sum over
$\ell$ and then the integration over $\theta$.
The sum diverges as $|g_{L}| \rightarrow 1$ ($x \rightarrow 0$
and $\theta \rightarrow 0$) but since the smallest $x$ we consider is
$x=.0001$ this is not a real problem - we just have to sum up more terms to
achieve a set accuracy. The integration over $\theta$ is
straightforwardly accomplished using an open Newton-Cotes algorithm.

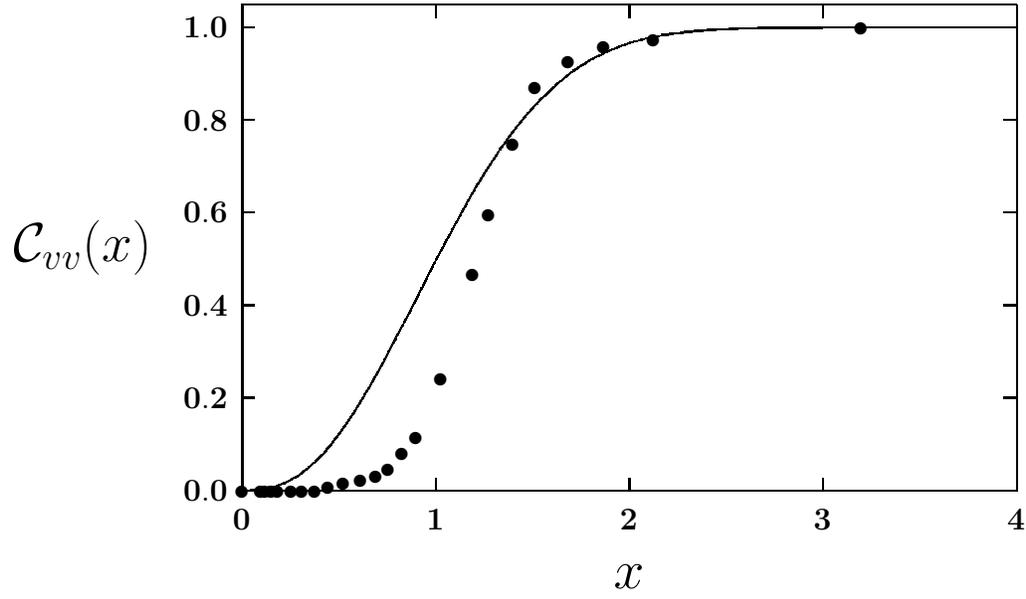
\begin{figure}
% GNUPLOT: LaTeX picture
\setlength{\unitlength}{0.240900pt}
\ifx\plotpoint\undefined\newsavebox{\plotpoint}\fi
\sbox{\plotpoint}{\rule[-0.200pt]{0.400pt}{0.400pt}}%
\begin{picture}(1500,900)(-150,0)
\font\gnuplot=cmr10 at 10pt
\gnuplot
\sbox{\plotpoint}{\rule[-0.200pt]{0.400pt}{0.400pt}}%
\put(220.0,113.0){\rule[-0.200pt]{0.400pt}{184.048pt}}
\put(220.0,113.0){\rule[-0.200pt]{4.818pt}{0.400pt}}
\put(198,113){\makebox(0,0)[r]{${\bf 0.0}$}}
\put(1416.0,113.0){\rule[-0.200pt]{4.818pt}{0.400pt}}
\put(220.0,259.0){\rule[-0.200pt]{4.818pt}{0.400pt}}
\put(198,259){\makebox(0,0)[r]{${\bf 0.2}$}}
\put(1416.0,259.0){\rule[-0.200pt]{4.818pt}{0.400pt}}
\put(220.0,404.0){\rule[-0.200pt]{4.818pt}{0.400pt}}
\put(198,404){\makebox(0,0)[r]{${\bf 0.4}$}}
\put(1416.0,404.0){\rule[-0.200pt]{4.818pt}{0.400pt}}
\put(220.0,550.0){\rule[-0.200pt]{4.818pt}{0.400pt}}
\put(198,550){\makebox(0,0)[r]{${\bf 0.6}$}}
\put(1416.0,550.0){\rule[-0.200pt]{4.818pt}{0.400pt}}
\put(220.0,695.0){\rule[-0.200pt]{4.818pt}{0.400pt}}
\put(198,695){\makebox(0,0)[r]{${\bf 0.8}$}}
\put(1416.0,695.0){\rule[-0.200pt]{4.818pt}{0.400pt}}
\put(220.0,841.0){\rule[-0.200pt]{4.818pt}{0.400pt}}
\put(198,841){\makebox(0,0)[r]{${\bf 1.0}$}}
\put(1416.0,841.0){\rule[-0.200pt]{4.818pt}{0.400pt}}
\put(220.0,113.0){\rule[-0.200pt]{0.400pt}{4.818pt}}
\put(220,68){\makebox(0,0){${\bf 0}$}}
\put(220.0,857.0){\rule[-0.200pt]{0.400pt}{4.818pt}}
\put(524.0,113.0){\rule[-0.200pt]{0.400pt}{4.818pt}}
\put(524,68){\makebox(0,0){${\bf 1}$}}
\put(524.0,857.0){\rule[-0.200pt]{0.400pt}{4.818pt}}
\put(828.0,113.0){\rule[-0.200pt]{0.400pt}{4.818pt}}
\put(828,68){\makebox(0,0){${\bf 2}$}}
\put(828.0,857.0){\rule[-0.200pt]{0.400pt}{4.818pt}}
\put(1132.0,113.0){\rule[-0.200pt]{0.400pt}{4.818pt}}
\put(1132,68){\makebox(0,0){${\bf 3}$}}
\put(1132.0,857.0){\rule[-0.200pt]{0.400pt}{4.818pt}}
\put(1436.0,113.0){\rule[-0.200pt]{0.400pt}{4.818pt}}
\put(1436,68){\makebox(0,0){${\bf 4}$}}
\put(1436.0,857.0){\rule[-0.200pt]{0.400pt}{4.818pt}}
\put(220.0,113.0){\rule[-0.200pt]{292.934pt}{0.400pt}}
\put(1436.0,113.0){\rule[-0.200pt]{0.400pt}{184.048pt}}
\put(220.0,877.0){\rule[-0.200pt]{292.934pt}{0.400pt}}
\put(-30,495){\makebox(0,0){\LARGE{${\cal C}_{vv}(x)$}}}
\put(828,-22){\makebox(0,0){\LARGE{$x$}}}
\put(220.0,113.0){\rule[-0.200pt]{0.400pt}{184.048pt}}
\put(221,113){\usebox{\plotpoint}}
\put(221,113){\usebox{\plotpoint}}
\put(221,113){\usebox{\plotpoint}}
\put(221,113){\usebox{\plotpoint}}
\put(221,113){\usebox{\plotpoint}}
\put(221,113){\usebox{\plotpoint}}
\put(221,113){\usebox{\plotpoint}}
\put(221,113){\usebox{\plotpoint}}
\put(234,112.67){\rule{0.482pt}{0.400pt}}
\multiput(234.00,112.17)(1.000,1.000){2}{\rule{0.241pt}{0.400pt}}
\put(221.0,113.0){\rule[-0.200pt]{3.132pt}{0.400pt}}
\put(246,113.67){\rule{1.445pt}{0.400pt}}
\multiput(246.00,113.17)(3.000,1.000){2}{\rule{0.723pt}{0.400pt}}
\put(252,114.67){\rule{1.445pt}{0.400pt}}
\multiput(252.00,114.17)(3.000,1.000){2}{\rule{0.723pt}{0.400pt}}
\put(258,116.17){\rule{1.500pt}{0.400pt}}
\multiput(258.00,115.17)(3.887,2.000){2}{\rule{0.750pt}{0.400pt}}
\put(265,117.67){\rule{1.686pt}{0.400pt}}
\multiput(265.00,117.17)(3.500,1.000){2}{\rule{0.843pt}{0.400pt}}
\multiput(272.00,119.61)(1.355,0.447){3}{\rule{1.033pt}{0.108pt}}
\multiput(272.00,118.17)(4.855,3.000){2}{\rule{0.517pt}{0.400pt}}
\multiput(279.00,122.61)(1.802,0.447){3}{\rule{1.300pt}{0.108pt}}
\multiput(279.00,121.17)(6.302,3.000){2}{\rule{0.650pt}{0.400pt}}
\multiput(288.00,125.59)(0.933,0.477){7}{\rule{0.820pt}{0.115pt}}
\multiput(288.00,124.17)(7.298,5.000){2}{\rule{0.410pt}{0.400pt}}
\multiput(297.00,130.59)(0.943,0.482){9}{\rule{0.833pt}{0.116pt}}
\multiput(297.00,129.17)(9.270,6.000){2}{\rule{0.417pt}{0.400pt}}
\multiput(308.00,136.58)(0.652,0.491){17}{\rule{0.620pt}{0.118pt}}
\multiput(308.00,135.17)(11.713,10.000){2}{\rule{0.310pt}{0.400pt}}
\multiput(321.00,146.58)(0.576,0.493){23}{\rule{0.562pt}{0.119pt}}
\multiput(321.00,145.17)(13.834,13.000){2}{\rule{0.281pt}{0.400pt}}
\multiput(336.00,159.58)(0.498,0.495){33}{\rule{0.500pt}{0.119pt}}
\multiput(336.00,158.17)(16.962,18.000){2}{\rule{0.250pt}{0.400pt}}
\multiput(354.58,177.00)(0.496,0.643){39}{\rule{0.119pt}{0.614pt}}
\multiput(353.17,177.00)(21.000,25.725){2}{\rule{0.400pt}{0.307pt}}
\multiput(375.58,204.00)(0.496,0.729){41}{\rule{0.120pt}{0.682pt}}
\multiput(374.17,204.00)(22.000,30.585){2}{\rule{0.400pt}{0.341pt}}
\multiput(397.58,236.00)(0.496,0.830){43}{\rule{0.120pt}{0.761pt}}
\multiput(396.17,236.00)(23.000,36.421){2}{\rule{0.400pt}{0.380pt}}
\multiput(420.58,274.00)(0.496,0.891){41}{\rule{0.120pt}{0.809pt}}
\multiput(419.17,274.00)(22.000,37.321){2}{\rule{0.400pt}{0.405pt}}
\multiput(442.58,313.00)(0.496,0.982){37}{\rule{0.119pt}{0.880pt}}
\multiput(441.17,313.00)(20.000,37.174){2}{\rule{0.400pt}{0.440pt}}
\multiput(462.58,352.00)(0.496,0.934){39}{\rule{0.119pt}{0.843pt}}
\multiput(461.17,352.00)(21.000,37.251){2}{\rule{0.400pt}{0.421pt}}
\multiput(483.58,391.00)(0.496,1.007){37}{\rule{0.119pt}{0.900pt}}
\multiput(482.17,391.00)(20.000,38.132){2}{\rule{0.400pt}{0.450pt}}
\multiput(503.58,431.00)(0.496,1.007){37}{\rule{0.119pt}{0.900pt}}
\multiput(502.17,431.00)(20.000,38.132){2}{\rule{0.400pt}{0.450pt}}
\multiput(523.58,471.00)(0.496,0.934){39}{\rule{0.119pt}{0.843pt}}
\multiput(522.17,471.00)(21.000,37.251){2}{\rule{0.400pt}{0.421pt}}
\multiput(544.58,510.00)(0.496,0.931){37}{\rule{0.119pt}{0.840pt}}
\multiput(543.17,510.00)(20.000,35.257){2}{\rule{0.400pt}{0.420pt}}
\multiput(564.58,547.00)(0.496,0.905){37}{\rule{0.119pt}{0.820pt}}
\multiput(563.17,547.00)(20.000,34.298){2}{\rule{0.400pt}{0.410pt}}
\multiput(584.58,583.00)(0.496,0.813){39}{\rule{0.119pt}{0.748pt}}
\multiput(583.17,583.00)(21.000,32.448){2}{\rule{0.400pt}{0.374pt}}
\multiput(605.58,617.00)(0.496,0.765){39}{\rule{0.119pt}{0.710pt}}
\multiput(604.17,617.00)(21.000,30.527){2}{\rule{0.400pt}{0.355pt}}
\multiput(626.58,649.00)(0.496,0.692){39}{\rule{0.119pt}{0.652pt}}
\multiput(625.17,649.00)(21.000,27.646){2}{\rule{0.400pt}{0.326pt}}
\multiput(647.58,678.00)(0.496,0.614){41}{\rule{0.120pt}{0.591pt}}
\multiput(646.17,678.00)(22.000,25.774){2}{\rule{0.400pt}{0.295pt}}
\multiput(669.58,705.00)(0.496,0.542){43}{\rule{0.120pt}{0.535pt}}
\multiput(668.17,705.00)(23.000,23.890){2}{\rule{0.400pt}{0.267pt}}
\multiput(692.00,730.58)(0.521,0.496){41}{\rule{0.518pt}{0.120pt}}
\multiput(692.00,729.17)(21.924,22.000){2}{\rule{0.259pt}{0.400pt}}
\multiput(715.00,752.58)(0.619,0.496){39}{\rule{0.595pt}{0.119pt}}
\multiput(715.00,751.17)(24.765,21.000){2}{\rule{0.298pt}{0.400pt}}
\multiput(741.00,773.58)(0.785,0.494){29}{\rule{0.725pt}{0.119pt}}
\multiput(741.00,772.17)(23.495,16.000){2}{\rule{0.363pt}{0.400pt}}
\multiput(766.00,789.58)(1.015,0.492){19}{\rule{0.900pt}{0.118pt}}
\multiput(766.00,788.17)(20.132,11.000){2}{\rule{0.450pt}{0.400pt}}
\multiput(788.00,800.58)(1.121,0.491){17}{\rule{0.980pt}{0.118pt}}
\multiput(788.00,799.17)(19.966,10.000){2}{\rule{0.490pt}{0.400pt}}
\multiput(810.00,810.59)(1.560,0.485){11}{\rule{1.300pt}{0.117pt}}
\multiput(810.00,809.17)(18.302,7.000){2}{\rule{0.650pt}{0.400pt}}
\multiput(831.00,817.59)(1.935,0.477){7}{\rule{1.540pt}{0.115pt}}
\multiput(831.00,816.17)(14.804,5.000){2}{\rule{0.770pt}{0.400pt}}
\multiput(849.00,822.60)(2.528,0.468){5}{\rule{1.900pt}{0.113pt}}
\multiput(849.00,821.17)(14.056,4.000){2}{\rule{0.950pt}{0.400pt}}
\multiput(867.00,826.61)(3.365,0.447){3}{\rule{2.233pt}{0.108pt}}
\multiput(867.00,825.17)(11.365,3.000){2}{\rule{1.117pt}{0.400pt}}
\put(883,829.17){\rule{3.300pt}{0.400pt}}
\multiput(883.00,828.17)(9.151,2.000){2}{\rule{1.650pt}{0.400pt}}
\put(899,831.17){\rule{3.300pt}{0.400pt}}
\multiput(899.00,830.17)(9.151,2.000){2}{\rule{1.650pt}{0.400pt}}
\put(915,833.17){\rule{3.100pt}{0.400pt}}
\multiput(915.00,832.17)(8.566,2.000){2}{\rule{1.550pt}{0.400pt}}
\put(930,834.67){\rule{3.373pt}{0.400pt}}
\multiput(930.00,834.17)(7.000,1.000){2}{\rule{1.686pt}{0.400pt}}
\put(944,835.67){\rule{3.373pt}{0.400pt}}
\multiput(944.00,835.17)(7.000,1.000){2}{\rule{1.686pt}{0.400pt}}
\put(958,836.67){\rule{3.373pt}{0.400pt}}
\multiput(958.00,836.17)(7.000,1.000){2}{\rule{1.686pt}{0.400pt}}
\put(236.0,114.0){\rule[-0.200pt]{2.409pt}{0.400pt}}
\put(985,837.67){\rule{3.132pt}{0.400pt}}
\multiput(985.00,837.17)(6.500,1.000){2}{\rule{1.566pt}{0.400pt}}
\put(972.0,838.0){\rule[-0.200pt]{3.132pt}{0.400pt}}
\put(1024,838.67){\rule{2.891pt}{0.400pt}}
\multiput(1024.00,838.17)(6.000,1.000){2}{\rule{1.445pt}{0.400pt}}
\put(998.0,839.0){\rule[-0.200pt]{6.263pt}{0.400pt}}
\put(1127,839.67){\rule{2.409pt}{0.400pt}}
\multiput(1127.00,839.17)(5.000,1.000){2}{\rule{1.204pt}{0.400pt}}
\put(1036.0,840.0){\rule[-0.200pt]{21.922pt}{0.400pt}}
\put(1137.0,841.0){\rule[-0.200pt]{72.029pt}{0.400pt}}
\put(220,113){\raisebox{-.8pt}{\makebox(0,0){$\bullet$}}}
\put(249,113){\raisebox{-.8pt}{\makebox(0,0){$\bullet$}}}
\put(256,113){\raisebox{-.8pt}{\makebox(0,0){$\bullet$}}}
\put(266,113){\raisebox{-.8pt}{\makebox(0,0){$\bullet$}}}
\put(276,113){\raisebox{-.8pt}{\makebox(0,0){$\bullet$}}}
\put(297,113){\raisebox{-.8pt}{\makebox(0,0){$\bullet$}}}
\put(314,113){\raisebox{-.8pt}{\makebox(0,0){$\bullet$}}}
\put(334,113){\raisebox{-.8pt}{\makebox(0,0){$\bullet$}}}
\put(355,119){\raisebox{-.8pt}{\makebox(0,0){$\bullet$}}}
\put(379,125){\raisebox{-.8pt}{\makebox(0,0){$\bullet$}}}
\put(406,131){\raisebox{-.8pt}{\makebox(0,0){$\bullet$}}}
\put(430,136){\raisebox{-.8pt}{\makebox(0,0){$\bullet$}}}
\put(449,148){\raisebox{-.8pt}{\makebox(0,0){$\bullet$}}}
\put(471,172){\raisebox{-.8pt}{\makebox(0,0){$\bullet$}}}
\put(493,197){\raisebox{-.8pt}{\makebox(0,0){$\bullet$}}}
\put(532,289){\raisebox{-.8pt}{\makebox(0,0){$\bullet$}}}
\put(582,453){\raisebox{-.8pt}{\makebox(0,0){$\bullet$}}}
\put(607,547){\raisebox{-.8pt}{\makebox(0,0){$\bullet$}}}
\put(645,659){\raisebox{-.8pt}{\makebox(0,0){$\bullet$}}}
\put(680,747){\raisebox{-.8pt}{\makebox(0,0){$\bullet$}}}
\put(732,788){\raisebox{-.8pt}{\makebox(0,0){$\bullet$}}}
\put(788,811){\raisebox{-.8pt}{\makebox(0,0){$\bullet$}}}
\put(866,823){\raisebox{-.8pt}{\makebox(0,0){$\bullet$}}}
\put(1192,841){\raisebox{-.8pt}{\makebox(0,0){$\bullet$}}}
\end{picture}
\vspace{.5 in}
\caption{The scaling form ${\cal C}_{vv}(x)$ for the vortex-vortex correlation
function. The solid curve is the result for the theory presented here.
The dots represent the simulation data \protect\cite{MONDELLO90}.}
\label{FIG:CVV}
\end{figure}

Using the relations (\ref{EQ:CVVSCAL}) and (\ref{EQ:CVASCAL}) we have
calculated the results for ${\cal C}_{vv}(x)$ and ${\cal C}_{va}(x)$, which
are presented in Figures \ref{FIG:CVV} and \ref{FIG:CVA} respectively.
As expected, ${\cal C}_{vv}(x)$ has a depletion zone at small-$x$, which
has a characteristic size $x \approx 1$ $(|{\bf r}| \approx L(t)$).
In contrast,  ${\cal C}_{va}(x)$ shows enhanced correlations in the same
range of $x$.
We present the results for ${\cal G}_{s}$ and ${\cal G}_{u}$ in
Figures \ref{FIG:GS} and \ref{FIG:GU}. Also shown in
Figure \ref{FIG:GS} is the result of the earlier theory \cite{LIU92b,LIU92a}
for ${\cal G}_{s}$
which displays the divergence resulting from neglecting fluctuations.
These four figures display the main results of this paper.

Also shown in these figures are the results of simulation \cite{MONDELLO90}.
The simulation data for the scaling forms for the
vortex-vortex correlation function  and the
vortex-antivortex correlation function were taken
directly from figures 8 and 9 of \cite{MONDELLO90}. Relations
(\ref{EQ:CVVSCAL}) and (\ref{EQ:CVASCAL}) were then used to calculate the
simulation results for ${\cal G}_{s}$ and ${\cal G}_{u}$.
As mentioned in the Introduction, experimental data \cite{NAGAYA95} also exist
for these quantities. The experimental data track the simulation results well,
but are more noisy and are therefore not shown here.
There is only one adjustable parameter in all these fits, which is the
(unknown) proportionality coefficient between the scaling length $L(t)$
used in
the theory and that used in simulations \cite{NORMALIZATION}.
We use this freedom to adjust the horizontal scale of the simulation data
in Figure \ref{FIG:CVA} to give the best match between theory and simulation
at intermediate to large-$x$. The same scaling factor is used to rescale the
simulation data in the other three figures. It is amusing to see that in
Figure \ref{FIG:GS} this rescaling allows the minima of the
simulation data and the theoretical curve for the earlier theory \cite{LIU92b}
to coincide. Figures \ref{FIG:CVV} and
\ref{FIG:CVA} show reasonable agreement between theory and simulation except
for the short distance behaviour of $C_{va}$.  This discrepancy is directly
related to the behavior of ${\cal G}_{s}$. Since we know that the
theoretical expression for ${\cal G}_{s}$ satisfies the sum rule
(\ref{EQ:SUMRULE}), the question is whether this sum rule is satisfied by
the simulation results. We have found that the data
presented in \cite{MONDELLO90} lead to the result
%
% SIMULATION RESULTS
%
\be
\label{EQ:SIMSUM}
- {A} \int ~d^{d}x~ {\cal G}_{s} (x) = 0.85 \pm 0.05,
\ee
which is less than the expected result of unity.
We are able to compute this quantity directly from the data contained in
figures 8 and 9 of \cite{MONDELLO90}, using the fact that in units of the
scaling length they use in these figures $n_{0}(t) = L^{-2}$ so $A = 1$.
The error estimate in (\ref{EQ:SIMSUM}) is due to the uncertainties in
reading the data, which is somewhat noisy, from the figures in
\cite{MONDELLO90}. What could account for this breakdown in the sum rule?
First, there is the possibility of a breakdown in scaling and
a violation of the scaling relations (\ref{EQ:GSSCAL}) and
(\ref{EQ:DENSCALREL}). This seems
incompatible with the simulation
results when viewed as a function of time.  A second, more likely, possibility
is that there are some missing vortex-antivortex pairs
in the simulation.  We speculate that there may be a problem keeping
track of annihilating pairs at short-scaled distances where they
may have a very high relative velocity \cite{CORE}.
 This may be the source of the
short-distance discrepancy in ${\cal C}_{va}$ between theory and simulation.
%
% CONCLUSION
%
\section{Conclusion}

Our work here has concentrated on the statistical properties of
point vortices in phase ordering systems.  The correlations among like-signed
vortices found here meet with our expectation that vortices
with the same charge
repel one another at short-scaled distances and that
screening of this repulsive interaction causes the correlations to fall
rapidly to zero at large-scaled distances.
The case of correlations between unlike-signed vortices seems straightforward
from the theoretical point of view.  Since these pairs are
attractive there is an increasing probability of finding pairs
on their way to annihilation as one goes to short-scaled distances.  The
simulation results seem at odds with this simple physical interpretation.
One argument around the monotonic behaviour of the vortex-antivortex
correlation function is that the annihilating pair is speeding up
in the late stage of annihilation and therefore the probability of
finding the pair separated by a short distance is commensurately decreased.
In recent calculations \cite{MAZENKO96b},
using fundamentally the same theory, one of us found that a mechanism
already exists in the theory to produce large vortex velocities.  These
large velocities are inferred from a power-law tail in the vortex
velocity probability distribution.  Bray \cite{BRAY96} has shown that
this tail
results from scaling arguments applied to the
the late stages of the  vortex-antivortex annihilation process.
Thus it appears that
this speeding up process is included in the present analysis.

In order to better understand the nature of the correlations between
vortex-antivortex pairs, it would be instructive to work out the
joint probability of having a vortex at position ${\bf r}$ with velocity
${\bf v}_{2}$ given that there is a vortex at the origin with velocity
${\bf v}_{1}$.  This calculation is under current investigation.

One of the remaining unresolved questions is:  Where are the missing
vortex-antivortex pairs in the simulations?  Further
progress is hindered until this discrepancy is understood.
%
% ACKNOWLEDGEMENTS
%
\acknowledgements
This work was supported in part by the MRSEC Program of the National Science
Foundation under Award Number DMR-9400379.
R.A.W. gratefully acknowledges support from the NSERC of Canada.
%
% APPENDIX A: CALCULATION OF THE 2-PT DISTRIBUTION
%
\appendix
\section{}

We derive an expression for the reduced probability distribution
%
% DEFINITION OF G
%
\be
G(\xi_{1},\xi_{2}) = \langle \delta[\M(1)] \delta[\M(2)] \prod_{\mu \nu}
         \delta[(\xi_{1})_{\mu}^{\nu} - \nabla_{\mu} m_{\nu} (1) ]
         \delta[(\xi_{2})_{\mu}^{\nu} - \nabla_{\mu} m_{\nu} (2) ] \rangle
\ee
appearing in the integral formula (\ref{EQ:GUINT})
for $\tilde{G}_{u} ({\bf r},t)$. We evaluate this
gaussian average for equal times  $t_{1} = t_{2} = t$. The $\delta$-functions
can be represented as integrals and one has
\be
%
% INTEGRAL REPRESENTATION OF G
%
G(\xi_{1},\xi_{2}) = \int
         \frac{d^{n}q_{1}}{(2 \pi)^{n}} \frac{d^{n}q_{2}}{(2 \pi)^{n}}
\prod_{\mu \nu} \frac{d (k_{1})_{\mu}^{\nu}}{2 \pi}
                \frac{d (k_{2})_{\mu}^{\nu}}{2 \pi}
          \mbox{ }
         \Gamma(\vec{q}_{1},\vec{q}_{2},k_{1},k_{2}) \mbox{ }
         \exp - i (k_{j})_{\mu}^{\nu} (\xi_{j})_{\mu}^{\nu}
\ee
where we have defined
%
% GAMMA DEFINITION
%
\be
\label{EQ:GAMMA}
\Gamma(\vec{q}_{1},\vec{q}_{2},k_{1},k_{2}) =
 \langle \exp - i [\vec{q}_{j} \cdot
\M (j) - (k_{j})_{\mu}^{\nu} \nabla_{\mu} m_{\nu} (j) ] \rangle.
\ee
In these formulae summation over the repeated indices $\mu$, $\nu$ and $j$ is
implied. The summation over $j$ is from $1$ to $2$, while the summation over
$\mu$ and $\nu$ is from $1$ to $n$, unless stated otherwise. Expression
(\ref{EQ:GAMMA}) is of the standard form for gaussian integrals
%
% STANDARD GAUSSIAN IDENTITY
%
\be
\label{EQ:HANDYID}
\langle \exp \int d \bar{1} \vec{H} (\bar{1}) \cdot \M (\bar{1}) \rangle
= \exp \frac{1}{2} \int d \bar{1} d \bar{2}  \vec{H} (\bar{1}) \cdot
\vec{H} (\bar{2}) C_{0}(\bar{1} \bar{2}),
\ee
so a straightforward calculation yields
%
% RESULT FOR GAMMA
%
\bea
\lefteqn{
2 \ln \Gamma(\vec{q}_{1},\vec{q}_{2},k_{1},k_{2}) =
- S_{0} \mbox{ } \sum_{j} \vec{q}_{j} \mbox{}^{2}
- S^{(2)} \mbox{ } \sum_{\mu \nu j}   [(k_{j})_{\mu}^{\nu}]^{2}
- 2  \mbox{ } C_{0} \mbox{ }  \vec{q}_{1} \cdot \vec{q}_{2}
} \nonumber  \\ & &
 + 2 \mbox{ }[ C_{0}' \mbox{ } ( (k_{1})_{\mu}^{\nu} q_{2}^{\nu}
\hat{r}_{\mu} -
 (k_{2})_{\mu}^{\nu}  q_{1}^{\nu} \hat{r}_{\mu} ) + (C_{0}'' -
\frac{C_{0}'}{r} ) \mbox{ }
(k_{1})_{\mu}^{\alpha} (k_{2})_{\nu}^{\alpha} \hat{r}_{\mu}
\hat{r}_{\nu} + \frac{C_{0}'}{r}
\mbox{ } (k_{1})_{\mu}^{\nu} (k_{2})_{\mu}^{\nu} ]
\eea
where primes indicate differentiation with respect to $r$.
This expression can be clarified if we introduce the
orthogonal matrices $\hat{M}_{\mu}^{\nu}$ where
%
% ORTHONORMAL SET
%
\be
\hat{M}_{\alpha}^{\mu} \hat{M}_{\alpha}^{\nu} =
\hat{M}_{\mu}^{\alpha} \hat{M}_{\nu}^{\alpha} = \delta_{\mu \nu}
\label{EQ:DEF1}
\ee
and
\be
\label{EQ:DEF2}
\hat{M}_{1}^{\mu} =  \hat{r}_{\mu},
\ee
and then transform to the new variables $(W_{j})_{\mu}^{\nu}$ defined by
%
% TRANSFORMATION
%
\be
\label{EQ:TRANSFORMATION}
(W_{j})_{\mu}^{\nu} = \hat{M}_{\alpha}^{\mu} (k_{j})_{\alpha}^{\nu}.
\ee
We then obtain
%
% SEPARATED FORM FOR GAMMA
%
\bea
\lefteqn{
\Gamma(\vec{q}_{1},\vec{q}_{2},W_{1},W_{2}) =
} \nonumber  \\ & &
\exp - \frac{1}{2} [
A(\vec{q}_{1},\vec{q}_{2}) + A_{L} ((\vec{W}_{1})_{L},(\vec{W}_{2})_{L}) +
A_{T} ((W_{1})_{T},(W_{2})_{T}) + A_{c} (\vec{q}_{1},\vec{q}_{2},
(\vec{W}_{1})_{L},(\vec{W}_{2})_{L})]
\label{EQ:GAMMA2}
\eea
where $(W_{j})_{L}^{\nu} = (W_{j})_{1}^{\nu}$ is the longitudinal part of
$W_{j}$ and $(W_{j})_{T}$
is a shorthand referring to the remaining transverse parts of $W_{j}$ (
$(W_{j})_{\mu}^{\nu}$ with $\mu > 1$ ).
With this choice of variables we notice that
$\Gamma$ can be factored into a transverse and a longitudinal piece. We see
that for our purposes
we are not required to be more explicit than (\ref{EQ:DEF1}) and
(\ref{EQ:DEF2}) in defining the $\hat{M}_{\mu}^{\nu}$. The quantities appearing
in the exponential in (\ref{EQ:GAMMA2}) are
%
% QUANTITIES IN THE EXPONENTIAL
%
\bea
A(\vec{q}_{1}, \vec{q}_{2}) & = & S_{0} \mbox{ } \sum_{j} \vec{q}_{j}
\mbox{}^{2} + 2 \mbox{ } C_{0} \mbox{ } \vec{q}_{1} \cdot \vec{q}_{2} \\
A_{L}((\vec{W}_{1})_{L}, (\vec{W}_{2})_{L}) & = & S^{(2)} \mbox{ } \sum_{j}
[(\vec{W}_{j})_{L}]^{2} - 2 \mbox{ }
C_{0}'' \mbox{ } (\vec{W}_{1})_{L} \cdot (\vec{W}_{2})_{L} \\
A_{T} ((W_{1})_{T},(W_{2})_{T}) & = &
-2 \frac{C_{0}'}{r} \mbox{ } \sum_{\mu = 2}
\sum_{\nu} (W_{1})_{\mu}^{\nu} (W_{2})_{\mu}^{\nu}
+ S^{(2)} \mbox{ } \sum_{\mu = 2} \sum_{\nu j}
 [(W_{j})_{\mu}^{\nu}]^{2}  \\
A_{c}(\vec{q}_{1},\vec{q}_{2},(\vec{W}_{1})_{L},(\vec{W}_{2})_{L})
& = & 2 \mbox{ }  C_{0}' \mbox{ }
[\vec{q}_{1} \cdot (\vec{W}_{2})_{L} - \vec{q}_{2} \cdot (\vec{W}_{1})_{L}].
\eea
One can integrate (\ref{EQ:GAMMA2}) over $\vec{q}_{1}$ and $\vec{q}_{2}$
to obtain
%
% GAMMA W
%
\bea
\Gamma(W_{1},W_{2}) & = & \int \frac{d^{n} q_{1}}{(2 \pi)^{n}}
\frac{d^{n} q_{2}}{(2 \pi)^{n}} \Gamma(\vec{q}_{1},\vec{q}_{2},W_{1},W_{2})
\nonumber \\
& = & \left( \frac{\gamma}{2 \pi S_{0}} \right)^{n} \exp -\frac{1}{2} [A_{L}'
((\vec{W}_{1})_{L},(\vec{W}_{2})_{L}) + A_{T}((W_{1})_{T},(W_{2})_{T})]
\eea
where we define
\be
A_{L}'((\vec{W}_{1})_{L},(\vec{W}_{2})_{L})  =
S_{L} \mbox{ } \sum_{j} [(\vec{W}_{j})_{L}]^{2} + 2 \mbox{ }
C_{L} \mbox{ } (\vec{W}_{1})_{L} \cdot (\vec{W}_{2})_{L}
\ee
with $S_{L}$ and $C_{L}$ given by (\ref{EQ:SL}) and (\ref{EQ:CL}) respectively.
Since the transformation (\ref{EQ:TRANSFORMATION}) is orthogonal
we have the simple result for $G$
%
% G DEFNITION
%
\be
\label{EQ:GFORM}
G(t_{1},t_{2}) =
\left( \frac{\gamma}{2 \pi S_{0}} \right)^{n} G_{L} ((\vec{t}_{1})_{L},
(\vec{t}_{2})_{L}) \mbox{ } G_{T} ((t_{1})_{T},(t_{2})_{T})
\ee
depending on the rotated variable
%
% T DEFINITION
%
\be
\label{EQ:T2}
(t_{j})_{\mu}^{\nu} = \hat{M}_{\beta}^{\mu} (\xi_{j})_{\beta}^{\nu}.
\ee
The longitudinal and transverse parts of $t$ are defined in analogy to those
of $W$. The functions $G_{L}$ and $G_{T}$ appearing in
(\ref{EQ:GFORM}) are explicitly given in (\ref{EQ:D1}) and (\ref{EQ:D2}).
%
% APPENDIX 2: CALCULATION OF J FOR N=D=2
%
\section{}

In this Appendix we compute the integral
%
% DEFINITION OF J
%
\be
J(\vec{s}_{1},\vec{s}_{2}) = \int d^{2} x d^{2} y
|s_{11} x_{2} - s_{12} x_{1}||s_{21} y_{2} - s_{22} y_{1}|
\exp - \frac{1}{2} [ \vec{x} \mbox{}^{2} + \vec{y} \mbox{}^{2} -
2 f_{T} \mbox{ } \vec{x} \cdot \vec{y} \mbox{ }]
\label{EQ:JDEFN}
\ee
which is needed to evaluate ${\cal G}_{u}({\bf r},t)$
 for $n=d=2$. The $j^{th}$
component of $\vec{s}_{i}$ is written $s_{ij}$. To rid ourselves of
the absolute values appearing in (\ref{EQ:JDEFN}) we make use of the identity
%
% USEFUL IDENTITY
%
\be
|x| = \int_{-\infty}^{+\infty} \frac{dz}{\sqrt{2 \pi}} \left( - \frac{1}{z}
\frac{\partial}{\partial z} \right) e^{-x^{2} z^{2} / 2}
\ee
and write
%
% NEW FORM FOR J
%
\be
\label{EQ:NEWJ}
J(\vec{s}_{1},\vec{s}_{2}) = \int_{-\infty}^{+\infty}
\frac{d z_{1}}{\sqrt{2 \pi}}\frac{d z_{2}}{\sqrt{2 \pi}} \frac{1}{z_{1} z_{2}}
\frac{\partial^{2}}{\partial z_{1} \partial z_{2}} \int d^{2} x
d^{2} y \exp - \frac{1}{2}
A(z_{1},z_{2},\vec{s}_{1},\vec{s}_{2},\vec{x},\vec{y})
\ee
with
%
% QUADRATIC FORM
%
\be
A(z_{1},z_{2},\vec{s}_{1},\vec{s}_{2},\vec{x},\vec{y}) = z_{1}^{2}
( s_{11} x_{2} - s_{12} x_{1})^{2} + z_{2}^{2} (s_{21}  y_{2}
- s_{22} y_{1})^{2}  + \vec{x} \mbox{}^{2} + \vec{y} \mbox{}^{2}
 - 2 f_{T} \mbox{ }
\vec{x} \cdot \vec{y}.
\ee
The integrations over $\vec{x}$ and $\vec{y}$ in (\ref{EQ:NEWJ}) are gaussian
and so can be readily, if somewhat tediously, performed. One has
%
% NEW2 FORM FOR J
%
\be
J(\vec{s}_{1},\vec{s}_{2}) = \int_{-\infty}^{+\infty} \frac{d z_{1}}
{\sqrt{2 \pi}}
\frac{d z_{2}}{\sqrt{2 \pi}} \frac{1}{z_{1} z_{2}}
\frac{\partial^{2}}{\partial z_{1} \partial z_{2}}
\frac{(2 \pi)^{2}}{\sqrt{D(z_{1},z_{2},\vec{s}_{1},\vec{s}_{2})}}
\label{EQ:NEWJ2}
\ee
where the determinant $D$ is given by
%
% DETERMINANT
%
\be
D = (1 + z_{2}^{2} \vec{s}_{2} \mbox{}^{2})(1 + z_{1}^{2} \vec{s}_{1}
\mbox{}^{2})
- f_{T}^{2} ( 2 + z_{1}^{2} \vec{s}_{1}
\mbox{}^{2} + z_{2}^{2} \vec{s}_{2} \mbox{}^{2} +
 z_{1}^{2} z_{2}^{2} [ \vec{s}_{1} \cdot \vec{s}_{2}]^{2}) + f_{T}^{4}.
\ee
The next step is to evaluate the derivatives with respect
to $z_{1}$ and $z_{2}$ appearing in (\ref{EQ:NEWJ2}).
This can be done straightforwardly, and one notices
that a change of variables allows one to write the integral in
(\ref{EQ:NEWJ2}) as a product of the amplitudes of
$\vec{s}_{1}$ and $\vec{s}_{2}$ and an integral whose only dependence on
$\vec{s}_{1}$ and $\vec{s}_{2}$  is through the dot-product form
%
% DEFINITION OF G
%
\be
g = f_{T}^{2} ( \hat{s}_{1} \cdot \hat{s}_{2} )^{2}.
\ee
Explicitly, one has
%
% SIMPLE FORM FOR J
%
\be
J(\vec{s}_{1},\vec{s}_{2})
= 2 \pi \gamma_{T}^{4} |\vec{s}_{1}||\vec{s}_{2}|  \tilde{J}(g)
\ee
with
%
% DEFINITION OF TILDE J
%
\be
\tilde{J}(g) = \int_{-\infty}^{+\infty} dy_{1} dy_{2} \frac{[1 + 2 g + (1-g)
(y_{1}^{2} + y_{2}^{2}) + (1-g)^{2} y_{1}^{2} y_{2}^{2} ]}{[ 1 + y_{1}^{2}
+ y_{2}^{2} + (1-g) y_{1}^{2} y_{2}^{2} ]^{5/2}}.
\ee
This seemingly complex integral for $\tilde{J}(g)$ is actually pleasingly
simple and after some manipulations one has
%
% SIMPLE FORM FOR TILDE J
%
\be
\tilde{J}(g) = 4 \sqrt{1 - g} + 4 \sqrt{g} \tan^{-1} \sqrt{\frac{g}{1-g}}.
\ee
%
% APPENDIX 3: SHORT DISTANCE BEHAVIOUR OF F ETC.
%
\section{}

To examine the small-$x$ behaviour of the defect correlation
functions we need to know the behaviour of $f$, $f_{T}$ and $f_{L}$ for small
$x$. In the theory we use here, which is discussed in detail in
\cite{MAZENKO96}, $f$ is analytic at short-scaled distances and we have
%
% F AT SMALL X
%
\be
\label{EQ:FEXPANSION}
f(x) = 1 - \alpha x^{2} + \beta x^{4} + \dots,
\ee
where $\alpha$ and $\beta$ are constants determined in the theory.
To evaluate $f_{T}$ we write $C_{T}$ (\ref{EQ:CT}) in terms of $f$, as a
function of the scaled length $x$:
%
% CT
%
\be
\label{EQ:CTSCAL}
C_{T} =  - \frac{S_{0}}{L^{2}} \frac{f'}{x} = - \frac{1}{2 d \alpha}
              \frac{f'}{x},
\ee
where we have used the result $S_{0}/L^{2}=1/2 d \alpha$ \cite{MAZENKO96}.
The result (\ref{EQ:CTSCAL}), together with the definition (\ref{EQ:ST}) for
$S_{T}$ and the expansion (\ref{EQ:FEXPANSION}) lead to
%
% SMALL X EXPANSION OF FT
%
\be
f_{T} \equiv \frac{C_{T}}{S_{T}} = 1 - \frac{2 \beta}{\alpha} x^{2}  + \dots
\ee
for small $x$.
We now examine $f_{L}$ in the scaling regime and use (\ref{EQ:SL}) and
(\ref{EQ:CL}) to write
%
% LONGITUDINAL NOTES
%
\be
S_{L} =  \frac{1}{d} - \frac{1}{2 d \alpha} \gamma^{2} (f')^{2}
\ee
and
\be
C_{L} =  - \frac{1}{2 d \alpha} [ f'' + f \gamma^{2} (f')^{2} ].
\ee
For small $x$ we again use (\ref{EQ:FEXPANSION}) and obtain
%
% FINAL FORM FOR f_{L}
%
\be
f_{L} \equiv \frac{C_{L}}{S_{L}} =  -1 + {\cal O} (x^{2}).
\ee
The key results here are $f_{T}\rightarrow 1$ while
$f_{L}\rightarrow -1$ when $x\rightarrow 0$.  This minus sign is ultimately
responsible
for the relation
%
% IMPORTANT REALTION
%
\be
\label{EQ:RR}
{\cal G}_{s}(0) = - {\cal G}_{u}(0)
\ee
between the signed and unsigned defect correlation functions at $x=0$
\cite{CAVEAT}.
%
% REFERENCES
%

%
% FIGURE 1: VORTEX-VORTEX CORRELATION FUNCTION
%
%
% FIGURE 2: VORTEX-ANTIVORTEX CORRELATION FUNCTION
%
\begin{figure}
% GNUPLOT: LaTeX picture
\setlength{\unitlength}{0.240900pt}
\ifx\plotpoint\undefined\newsavebox{\plotpoint}\fi
\sbox{\plotpoint}{\rule[-0.200pt]{0.400pt}{0.400pt}}%
\begin{picture}(1500,900)(-150,0)
\font\gnuplot=cmr10 at 10pt
\gnuplot
\sbox{\plotpoint}{\rule[-0.200pt]{0.400pt}{0.400pt}}%
\put(220.0,113.0){\rule[-0.200pt]{0.400pt}{184.048pt}}
\put(220.0,113.0){\rule[-0.200pt]{4.818pt}{0.400pt}}
\put(198,113){\makebox(0,0)[r]{${\bf 0.0}$}}
\put(1416.0,113.0){\rule[-0.200pt]{4.818pt}{0.400pt}}
\put(220.0,368.0){\rule[-0.200pt]{4.818pt}{0.400pt}}
\put(198,368){\makebox(0,0)[r]{${\bf 1.0}$}}
\put(1416.0,368.0){\rule[-0.200pt]{4.818pt}{0.400pt}}
\put(220.0,622.0){\rule[-0.200pt]{4.818pt}{0.400pt}}
\put(198,622){\makebox(0,0)[r]{${\bf 2.0}$}}
\put(1416.0,622.0){\rule[-0.200pt]{4.818pt}{0.400pt}}
\put(220.0,877.0){\rule[-0.200pt]{4.818pt}{0.400pt}}
\put(198,877){\makebox(0,0)[r]{${\bf 3.0}$}}
\put(1416.0,877.0){\rule[-0.200pt]{4.818pt}{0.400pt}}
\put(220.0,113.0){\rule[-0.200pt]{0.400pt}{4.818pt}}
\put(220,68){\makebox(0,0){${\bf 0}$}}
\put(220.0,857.0){\rule[-0.200pt]{0.400pt}{4.818pt}}
\put(524.0,113.0){\rule[-0.200pt]{0.400pt}{4.818pt}}
\put(524,68){\makebox(0,0){${\bf 1}$}}
\put(524.0,857.0){\rule[-0.200pt]{0.400pt}{4.818pt}}
\put(828.0,113.0){\rule[-0.200pt]{0.400pt}{4.818pt}}
\put(828,68){\makebox(0,0){${\bf 2}$}}
\put(828.0,857.0){\rule[-0.200pt]{0.400pt}{4.818pt}}
\put(1132.0,113.0){\rule[-0.200pt]{0.400pt}{4.818pt}}
\put(1132,68){\makebox(0,0){${\bf 3}$}}
\put(1132.0,857.0){\rule[-0.200pt]{0.400pt}{4.818pt}}
\put(1436.0,113.0){\rule[-0.200pt]{0.400pt}{4.818pt}}
\put(1436,68){\makebox(0,0){${\bf 4}$}}
\put(1436.0,857.0){\rule[-0.200pt]{0.400pt}{4.818pt}}
\put(220.0,113.0){\rule[-0.200pt]{292.934pt}{0.400pt}}
\put(1436.0,113.0){\rule[-0.200pt]{0.400pt}{184.048pt}}
\put(220.0,877.0){\rule[-0.200pt]{292.934pt}{0.400pt}}
\put(-30,495){\makebox(0,0){\LARGE{${\cal C}_{va}(x)$}}}
\put(828,-22){\makebox(0,0){\LARGE{$x$}}}
\put(220.0,113.0){\rule[-0.200pt]{0.400pt}{184.048pt}}
\put(221,868){\usebox{\plotpoint}}
\put(221,863.67){\rule{0.241pt}{0.400pt}}
\multiput(221.00,864.17)(0.500,-1.000){2}{\rule{0.120pt}{0.400pt}}
\put(221.0,865.0){\rule[-0.200pt]{0.400pt}{0.723pt}}
\put(222,864){\usebox{\plotpoint}}
\put(222.0,862.0){\rule[-0.200pt]{0.400pt}{0.482pt}}
\put(222.0,862.0){\usebox{\plotpoint}}
\put(223,859.67){\rule{0.241pt}{0.400pt}}
\multiput(223.00,860.17)(0.500,-1.000){2}{\rule{0.120pt}{0.400pt}}
\put(223.0,861.0){\usebox{\plotpoint}}
\put(224.0,859.0){\usebox{\plotpoint}}
\put(224.0,859.0){\usebox{\plotpoint}}
\put(225,856.67){\rule{0.241pt}{0.400pt}}
\multiput(225.00,857.17)(0.500,-1.000){2}{\rule{0.120pt}{0.400pt}}
\put(225.67,855){\rule{0.400pt}{0.482pt}}
\multiput(225.17,856.00)(1.000,-1.000){2}{\rule{0.400pt}{0.241pt}}
\put(227,853.67){\rule{0.241pt}{0.400pt}}
\multiput(227.00,854.17)(0.500,-1.000){2}{\rule{0.120pt}{0.400pt}}
\put(228,852.67){\rule{0.241pt}{0.400pt}}
\multiput(228.00,853.17)(0.500,-1.000){2}{\rule{0.120pt}{0.400pt}}
\put(228.67,851){\rule{0.400pt}{0.482pt}}
\multiput(228.17,852.00)(1.000,-1.000){2}{\rule{0.400pt}{0.241pt}}
\put(230,849.67){\rule{0.482pt}{0.400pt}}
\multiput(230.00,850.17)(1.000,-1.000){2}{\rule{0.241pt}{0.400pt}}
\put(232,848.17){\rule{0.482pt}{0.400pt}}
\multiput(232.00,849.17)(1.000,-2.000){2}{\rule{0.241pt}{0.400pt}}
\put(234.17,845){\rule{0.400pt}{0.700pt}}
\multiput(233.17,846.55)(2.000,-1.547){2}{\rule{0.400pt}{0.350pt}}
\multiput(236.00,843.95)(0.462,-0.447){3}{\rule{0.500pt}{0.108pt}}
\multiput(236.00,844.17)(1.962,-3.000){2}{\rule{0.250pt}{0.400pt}}
\multiput(239.00,840.95)(0.462,-0.447){3}{\rule{0.500pt}{0.108pt}}
\multiput(239.00,841.17)(1.962,-3.000){2}{\rule{0.250pt}{0.400pt}}
\multiput(242.60,836.51)(0.468,-0.627){5}{\rule{0.113pt}{0.600pt}}
\multiput(241.17,837.75)(4.000,-3.755){2}{\rule{0.400pt}{0.300pt}}
\multiput(246.00,832.93)(0.491,-0.482){9}{\rule{0.500pt}{0.116pt}}
\multiput(246.00,833.17)(4.962,-6.000){2}{\rule{0.250pt}{0.400pt}}
\multiput(252.59,825.37)(0.482,-0.671){9}{\rule{0.116pt}{0.633pt}}
\multiput(251.17,826.69)(6.000,-6.685){2}{\rule{0.400pt}{0.317pt}}
\multiput(258.00,818.93)(0.492,-0.485){11}{\rule{0.500pt}{0.117pt}}
\multiput(258.00,819.17)(5.962,-7.000){2}{\rule{0.250pt}{0.400pt}}
\multiput(265.59,810.45)(0.485,-0.645){11}{\rule{0.117pt}{0.614pt}}
\multiput(264.17,811.73)(7.000,-7.725){2}{\rule{0.400pt}{0.307pt}}
\multiput(272.59,801.45)(0.485,-0.645){11}{\rule{0.117pt}{0.614pt}}
\multiput(271.17,802.73)(7.000,-7.725){2}{\rule{0.400pt}{0.307pt}}
\multiput(279.59,792.37)(0.489,-0.669){15}{\rule{0.118pt}{0.633pt}}
\multiput(278.17,793.69)(9.000,-10.685){2}{\rule{0.400pt}{0.317pt}}
\multiput(288.59,780.37)(0.489,-0.669){15}{\rule{0.118pt}{0.633pt}}
\multiput(287.17,781.69)(9.000,-10.685){2}{\rule{0.400pt}{0.317pt}}
\multiput(297.58,768.17)(0.492,-0.732){19}{\rule{0.118pt}{0.682pt}}
\multiput(296.17,769.58)(11.000,-14.585){2}{\rule{0.400pt}{0.341pt}}
\multiput(308.58,752.29)(0.493,-0.695){23}{\rule{0.119pt}{0.654pt}}
\multiput(307.17,753.64)(13.000,-16.643){2}{\rule{0.400pt}{0.327pt}}
\multiput(321.58,734.15)(0.494,-0.737){27}{\rule{0.119pt}{0.687pt}}
\multiput(320.17,735.57)(15.000,-20.575){2}{\rule{0.400pt}{0.343pt}}
\multiput(336.58,712.19)(0.495,-0.725){33}{\rule{0.119pt}{0.678pt}}
\multiput(335.17,713.59)(18.000,-24.593){2}{\rule{0.400pt}{0.339pt}}
\multiput(354.58,686.13)(0.496,-0.740){39}{\rule{0.119pt}{0.690pt}}
\multiput(353.17,687.57)(21.000,-29.567){2}{\rule{0.400pt}{0.345pt}}
\multiput(375.58,655.32)(0.496,-0.683){41}{\rule{0.120pt}{0.645pt}}
\multiput(374.17,656.66)(22.000,-28.660){2}{\rule{0.400pt}{0.323pt}}
\multiput(397.58,625.35)(0.496,-0.675){43}{\rule{0.120pt}{0.639pt}}
\multiput(396.17,626.67)(23.000,-29.673){2}{\rule{0.400pt}{0.320pt}}
\multiput(420.58,594.55)(0.496,-0.614){41}{\rule{0.120pt}{0.591pt}}
\multiput(419.17,595.77)(22.000,-25.774){2}{\rule{0.400pt}{0.295pt}}
\multiput(442.58,567.59)(0.496,-0.600){37}{\rule{0.119pt}{0.580pt}}
\multiput(441.17,568.80)(20.000,-22.796){2}{\rule{0.400pt}{0.290pt}}
\multiput(462.58,543.85)(0.496,-0.522){39}{\rule{0.119pt}{0.519pt}}
\multiput(461.17,544.92)(21.000,-20.923){2}{\rule{0.400pt}{0.260pt}}
\multiput(483.00,522.92)(0.525,-0.495){35}{\rule{0.521pt}{0.119pt}}
\multiput(483.00,523.17)(18.919,-19.000){2}{\rule{0.261pt}{0.400pt}}
\multiput(503.00,503.92)(0.554,-0.495){33}{\rule{0.544pt}{0.119pt}}
\multiput(503.00,504.17)(18.870,-18.000){2}{\rule{0.272pt}{0.400pt}}
\multiput(523.00,485.92)(0.657,-0.494){29}{\rule{0.625pt}{0.119pt}}
\multiput(523.00,486.17)(19.703,-16.000){2}{\rule{0.313pt}{0.400pt}}
\multiput(544.00,469.92)(0.668,-0.494){27}{\rule{0.633pt}{0.119pt}}
\multiput(544.00,470.17)(18.685,-15.000){2}{\rule{0.317pt}{0.400pt}}
\multiput(564.00,454.92)(0.774,-0.493){23}{\rule{0.715pt}{0.119pt}}
\multiput(564.00,455.17)(18.515,-13.000){2}{\rule{0.358pt}{0.400pt}}
\multiput(584.00,441.92)(0.967,-0.492){19}{\rule{0.864pt}{0.118pt}}
\multiput(584.00,442.17)(19.207,-11.000){2}{\rule{0.432pt}{0.400pt}}
\multiput(605.00,430.92)(1.069,-0.491){17}{\rule{0.940pt}{0.118pt}}
\multiput(605.00,431.17)(19.049,-10.000){2}{\rule{0.470pt}{0.400pt}}
\multiput(626.00,420.93)(1.194,-0.489){15}{\rule{1.033pt}{0.118pt}}
\multiput(626.00,421.17)(18.855,-9.000){2}{\rule{0.517pt}{0.400pt}}
\multiput(647.00,411.93)(1.418,-0.488){13}{\rule{1.200pt}{0.117pt}}
\multiput(647.00,412.17)(19.509,-8.000){2}{\rule{0.600pt}{0.400pt}}
\multiput(669.00,403.93)(1.713,-0.485){11}{\rule{1.414pt}{0.117pt}}
\multiput(669.00,404.17)(20.065,-7.000){2}{\rule{0.707pt}{0.400pt}}
\multiput(692.00,396.93)(2.027,-0.482){9}{\rule{1.633pt}{0.116pt}}
\multiput(692.00,397.17)(19.610,-6.000){2}{\rule{0.817pt}{0.400pt}}
\multiput(715.00,390.93)(2.825,-0.477){7}{\rule{2.180pt}{0.115pt}}
\multiput(715.00,391.17)(21.475,-5.000){2}{\rule{1.090pt}{0.400pt}}
\multiput(741.00,385.94)(3.552,-0.468){5}{\rule{2.600pt}{0.113pt}}
\multiput(741.00,386.17)(19.604,-4.000){2}{\rule{1.300pt}{0.400pt}}
\multiput(766.00,381.95)(4.704,-0.447){3}{\rule{3.033pt}{0.108pt}}
\multiput(766.00,382.17)(15.704,-3.000){2}{\rule{1.517pt}{0.400pt}}
\multiput(788.00,378.95)(4.704,-0.447){3}{\rule{3.033pt}{0.108pt}}
\multiput(788.00,379.17)(15.704,-3.000){2}{\rule{1.517pt}{0.400pt}}
\put(810,375.17){\rule{4.300pt}{0.400pt}}
\multiput(810.00,376.17)(12.075,-2.000){2}{\rule{2.150pt}{0.400pt}}
\put(831,373.67){\rule{4.336pt}{0.400pt}}
\multiput(831.00,374.17)(9.000,-1.000){2}{\rule{2.168pt}{0.400pt}}
\put(849,372.67){\rule{4.336pt}{0.400pt}}
\multiput(849.00,373.17)(9.000,-1.000){2}{\rule{2.168pt}{0.400pt}}
\put(867,371.67){\rule{3.854pt}{0.400pt}}
\multiput(867.00,372.17)(8.000,-1.000){2}{\rule{1.927pt}{0.400pt}}
\put(883,370.67){\rule{3.854pt}{0.400pt}}
\multiput(883.00,371.17)(8.000,-1.000){2}{\rule{1.927pt}{0.400pt}}
\put(899,369.67){\rule{3.854pt}{0.400pt}}
\multiput(899.00,370.17)(8.000,-1.000){2}{\rule{1.927pt}{0.400pt}}
\put(225.0,858.0){\usebox{\plotpoint}}
\put(944,368.67){\rule{3.373pt}{0.400pt}}
\multiput(944.00,369.17)(7.000,-1.000){2}{\rule{1.686pt}{0.400pt}}
\put(915.0,370.0){\rule[-0.200pt]{6.986pt}{0.400pt}}
\put(998,367.67){\rule{3.132pt}{0.400pt}}
\multiput(998.00,368.17)(6.500,-1.000){2}{\rule{1.566pt}{0.400pt}}
\put(958.0,369.0){\rule[-0.200pt]{9.636pt}{0.400pt}}
\put(1011.0,368.0){\rule[-0.200pt]{102.382pt}{0.400pt}}
\put(220,113){\raisebox{-.8pt}{\makebox(0,0){$\bullet$}}}
\put(249,178){\raisebox{-.8pt}{\makebox(0,0){$\bullet$}}}
\put(256,211){\raisebox{-.8pt}{\makebox(0,0){$\bullet$}}}
\put(266,257){\raisebox{-.8pt}{\makebox(0,0){$\bullet$}}}
\put(276,296){\raisebox{-.8pt}{\makebox(0,0){$\bullet$}}}
\put(297,342){\raisebox{-.8pt}{\makebox(0,0){$\bullet$}}}
\put(314,394){\raisebox{-.8pt}{\makebox(0,0){$\bullet$}}}
\put(334,453){\raisebox{-.8pt}{\makebox(0,0){$\bullet$}}}
\put(355,505){\raisebox{-.8pt}{\makebox(0,0){$\bullet$}}}
\put(379,531){\raisebox{-.8pt}{\makebox(0,0){$\bullet$}}}
\put(406,544){\raisebox{-.8pt}{\makebox(0,0){$\bullet$}}}
\put(430,551){\raisebox{-.8pt}{\makebox(0,0){$\bullet$}}}
\put(449,545){\raisebox{-.8pt}{\makebox(0,0){$\bullet$}}}
\put(471,537){\raisebox{-.8pt}{\makebox(0,0){$\bullet$}}}
\put(493,521){\raisebox{-.8pt}{\makebox(0,0){$\bullet$}}}
\put(532,485){\raisebox{-.8pt}{\makebox(0,0){$\bullet$}}}
\put(582,446){\raisebox{-.8pt}{\makebox(0,0){$\bullet$}}}
\put(607,426){\raisebox{-.8pt}{\makebox(0,0){$\bullet$}}}
\put(645,407){\raisebox{-.8pt}{\makebox(0,0){$\bullet$}}}
\put(680,394){\raisebox{-.8pt}{\makebox(0,0){$\bullet$}}}
\put(732,381){\raisebox{-.8pt}{\makebox(0,0){$\bullet$}}}
\put(788,374){\raisebox{-.8pt}{\makebox(0,0){$\bullet$}}}
\put(866,368){\raisebox{-.8pt}{\makebox(0,0){$\bullet$}}}
\put(1192,368){\raisebox{-.8pt}{\makebox(0,0){$\bullet$}}}
\end{picture}
\vspace{.5 in}
\caption{The scaling form ${\cal C}_{va}(x)$ for the vortex-antivortex
correlation function. The solid curve is the result for the theory presented
here. The dots represent the simulation data \protect\cite{MONDELLO90}.}
\label{FIG:CVA}
\end{figure}
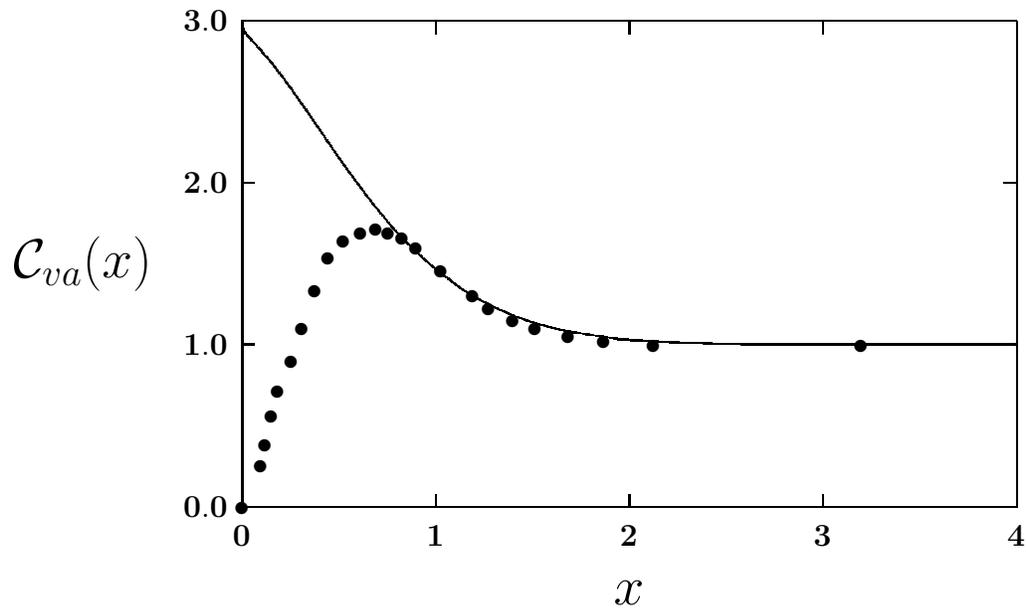
%
% FIGURE 3: SIGNED DEFECT-DEFECT CORRELATIONS
%
\begin{figure}
% GNUPLOT: LaTeX picture
\setlength{\unitlength}{0.240900pt}
\ifx\plotpoint\undefined\newsavebox{\plotpoint}\fi
\sbox{\plotpoint}{\rule[-0.200pt]{0.400pt}{0.400pt}}%
\begin{picture}(1500,900)(-150,0)
\font\gnuplot=cmr10 at 10pt
\gnuplot
\sbox{\plotpoint}{\rule[-0.200pt]{0.400pt}{0.400pt}}%
\put(220.0,113.0){\rule[-0.200pt]{0.400pt}{184.048pt}}
\put(220.0,149.0){\rule[-0.200pt]{4.818pt}{0.400pt}}
\put(198,149){\makebox(0,0)[r]{${\bf -1.5}$}}
\put(1416.0,149.0){\rule[-0.200pt]{4.818pt}{0.400pt}}
\put(220.0,331.0){\rule[-0.200pt]{4.818pt}{0.400pt}}
\put(198,331){\makebox(0,0)[r]{${\bf -1.0}$}}
\put(1416.0,331.0){\rule[-0.200pt]{4.818pt}{0.400pt}}
\put(220.0,513.0){\rule[-0.200pt]{4.818pt}{0.400pt}}
\put(198,513){\makebox(0,0)[r]{${\bf -0.5}$}}
\put(1416.0,513.0){\rule[-0.200pt]{4.818pt}{0.400pt}}
\put(220.0,695.0){\rule[-0.200pt]{4.818pt}{0.400pt}}
\put(198,695){\makebox(0,0)[r]{${\bf 0.0}$}}
\put(1416.0,695.0){\rule[-0.200pt]{4.818pt}{0.400pt}}
\put(220.0,877.0){\rule[-0.200pt]{4.818pt}{0.400pt}}
\put(198,877){\makebox(0,0)[r]{${\bf 0.5}$}}
\put(1416.0,877.0){\rule[-0.200pt]{4.818pt}{0.400pt}}
\put(220.0,113.0){\rule[-0.200pt]{0.400pt}{4.818pt}}
\put(220,68){\makebox(0,0){${\bf 0}$}}
\put(220.0,857.0){\rule[-0.200pt]{0.400pt}{4.818pt}}
\put(594.0,113.0){\rule[-0.200pt]{0.400pt}{4.818pt}}
\put(594,68){\makebox(0,0){${\bf 1}$}}
\put(594.0,857.0){\rule[-0.200pt]{0.400pt}{4.818pt}}
\put(968.0,113.0){\rule[-0.200pt]{0.400pt}{4.818pt}}
\put(968,68){\makebox(0,0){${\bf 2}$}}
\put(968.0,857.0){\rule[-0.200pt]{0.400pt}{4.818pt}}
\put(1342.0,113.0){\rule[-0.200pt]{0.400pt}{4.818pt}}
\put(1342,68){\makebox(0,0){${\bf 3}$}}
\put(1342.0,857.0){\rule[-0.200pt]{0.400pt}{4.818pt}}
\put(220.0,113.0){\rule[-0.200pt]{292.934pt}{0.400pt}}
\put(1436.0,113.0){\rule[-0.200pt]{0.400pt}{184.048pt}}
\put(220.0,877.0){\rule[-0.200pt]{292.934pt}{0.400pt}}
\put(-30,495){\makebox(0,0){\LARGE{${\cal G}_{s}(x)$}}}
\put(828,-22){\makebox(0,0){\LARGE{$x$}}}
\put(220.0,113.0){\rule[-0.200pt]{0.400pt}{184.048pt}}
\put(221,156){\usebox{\plotpoint}}
\put(221,156){\usebox{\plotpoint}}
\put(221.0,156.0){\rule[-0.200pt]{0.400pt}{0.482pt}}
\put(221.0,158.0){\usebox{\plotpoint}}
\put(222,158.67){\rule{0.241pt}{0.400pt}}
\multiput(222.00,158.17)(0.500,1.000){2}{\rule{0.120pt}{0.400pt}}
\put(222.0,158.0){\usebox{\plotpoint}}
\put(223,160){\usebox{\plotpoint}}
\put(223,160){\usebox{\plotpoint}}
\put(223,159.67){\rule{0.241pt}{0.400pt}}
\multiput(223.00,159.17)(0.500,1.000){2}{\rule{0.120pt}{0.400pt}}
\put(224,161){\usebox{\plotpoint}}
\put(224,160.67){\rule{0.241pt}{0.400pt}}
\multiput(224.00,160.17)(0.500,1.000){2}{\rule{0.120pt}{0.400pt}}
\put(225,161.67){\rule{0.241pt}{0.400pt}}
\multiput(225.00,161.17)(0.500,1.000){2}{\rule{0.120pt}{0.400pt}}
\put(227,162.67){\rule{0.241pt}{0.400pt}}
\multiput(227.00,162.17)(0.500,1.000){2}{\rule{0.120pt}{0.400pt}}
\put(228,163.67){\rule{0.241pt}{0.400pt}}
\multiput(228.00,163.17)(0.500,1.000){2}{\rule{0.120pt}{0.400pt}}
\put(229,164.67){\rule{0.241pt}{0.400pt}}
\multiput(229.00,164.17)(0.500,1.000){2}{\rule{0.120pt}{0.400pt}}
\put(230,165.67){\rule{0.241pt}{0.400pt}}
\multiput(230.00,165.17)(0.500,1.000){2}{\rule{0.120pt}{0.400pt}}
\put(231,166.67){\rule{0.482pt}{0.400pt}}
\multiput(231.00,166.17)(1.000,1.000){2}{\rule{0.241pt}{0.400pt}}
\put(233,167.67){\rule{0.482pt}{0.400pt}}
\multiput(233.00,167.17)(1.000,1.000){2}{\rule{0.241pt}{0.400pt}}
\put(235,168.67){\rule{0.482pt}{0.400pt}}
\multiput(235.00,168.17)(1.000,1.000){2}{\rule{0.241pt}{0.400pt}}
\put(237,170.17){\rule{0.700pt}{0.400pt}}
\multiput(237.00,169.17)(1.547,2.000){2}{\rule{0.350pt}{0.400pt}}
\put(240,172.17){\rule{0.700pt}{0.400pt}}
\multiput(240.00,171.17)(1.547,2.000){2}{\rule{0.350pt}{0.400pt}}
\multiput(243.00,174.61)(0.685,0.447){3}{\rule{0.633pt}{0.108pt}}
\multiput(243.00,173.17)(2.685,3.000){2}{\rule{0.317pt}{0.400pt}}
\multiput(247.00,177.61)(0.909,0.447){3}{\rule{0.767pt}{0.108pt}}
\multiput(247.00,176.17)(3.409,3.000){2}{\rule{0.383pt}{0.400pt}}
\multiput(252.00,180.59)(0.710,0.477){7}{\rule{0.660pt}{0.115pt}}
\multiput(252.00,179.17)(5.630,5.000){2}{\rule{0.330pt}{0.400pt}}
\multiput(259.00,185.59)(0.671,0.482){9}{\rule{0.633pt}{0.116pt}}
\multiput(259.00,184.17)(6.685,6.000){2}{\rule{0.317pt}{0.400pt}}
\multiput(267.00,191.59)(0.821,0.477){7}{\rule{0.740pt}{0.115pt}}
\multiput(267.00,190.17)(6.464,5.000){2}{\rule{0.370pt}{0.400pt}}
\multiput(275.00,196.59)(0.645,0.485){11}{\rule{0.614pt}{0.117pt}}
\multiput(275.00,195.17)(7.725,7.000){2}{\rule{0.307pt}{0.400pt}}
\multiput(284.00,203.59)(0.645,0.485){11}{\rule{0.614pt}{0.117pt}}
\multiput(284.00,202.17)(7.725,7.000){2}{\rule{0.307pt}{0.400pt}}
\multiput(293.00,210.59)(0.611,0.489){15}{\rule{0.589pt}{0.118pt}}
\multiput(293.00,209.17)(9.778,9.000){2}{\rule{0.294pt}{0.400pt}}
\multiput(304.00,219.58)(0.496,0.492){19}{\rule{0.500pt}{0.118pt}}
\multiput(304.00,218.17)(9.962,11.000){2}{\rule{0.250pt}{0.400pt}}
\multiput(315.00,230.58)(0.582,0.492){21}{\rule{0.567pt}{0.119pt}}
\multiput(315.00,229.17)(12.824,12.000){2}{\rule{0.283pt}{0.400pt}}
\multiput(329.58,242.00)(0.494,0.531){27}{\rule{0.119pt}{0.527pt}}
\multiput(328.17,242.00)(15.000,14.907){2}{\rule{0.400pt}{0.263pt}}
\multiput(344.00,258.58)(0.498,0.495){35}{\rule{0.500pt}{0.119pt}}
\multiput(344.00,257.17)(17.962,19.000){2}{\rule{0.250pt}{0.400pt}}
\multiput(363.00,277.58)(0.498,0.496){41}{\rule{0.500pt}{0.120pt}}
\multiput(363.00,276.17)(20.962,22.000){2}{\rule{0.250pt}{0.400pt}}
\multiput(385.58,299.00)(0.497,0.557){49}{\rule{0.120pt}{0.546pt}}
\multiput(384.17,299.00)(26.000,27.866){2}{\rule{0.400pt}{0.273pt}}
\multiput(411.58,328.00)(0.497,0.555){51}{\rule{0.120pt}{0.544pt}}
\multiput(410.17,328.00)(27.000,28.870){2}{\rule{0.400pt}{0.272pt}}
\multiput(438.58,358.00)(0.497,0.571){53}{\rule{0.120pt}{0.557pt}}
\multiput(437.17,358.00)(28.000,30.844){2}{\rule{0.400pt}{0.279pt}}
\multiput(466.58,390.00)(0.497,0.517){51}{\rule{0.120pt}{0.515pt}}
\multiput(465.17,390.00)(27.000,26.931){2}{\rule{0.400pt}{0.257pt}}
\multiput(493.58,418.00)(0.497,0.539){47}{\rule{0.120pt}{0.532pt}}
\multiput(492.17,418.00)(25.000,25.896){2}{\rule{0.400pt}{0.266pt}}
\multiput(518.58,445.00)(0.497,0.519){47}{\rule{0.120pt}{0.516pt}}
\multiput(517.17,445.00)(25.000,24.929){2}{\rule{0.400pt}{0.258pt}}
\multiput(543.00,471.58)(0.519,0.496){45}{\rule{0.517pt}{0.120pt}}
\multiput(543.00,470.17)(23.928,24.000){2}{\rule{0.258pt}{0.400pt}}
\multiput(568.00,495.58)(0.542,0.496){43}{\rule{0.535pt}{0.120pt}}
\multiput(568.00,494.17)(23.890,23.000){2}{\rule{0.267pt}{0.400pt}}
\multiput(593.00,518.58)(0.595,0.496){39}{\rule{0.576pt}{0.119pt}}
\multiput(593.00,517.17)(23.804,21.000){2}{\rule{0.288pt}{0.400pt}}
\multiput(618.00,539.58)(0.659,0.495){35}{\rule{0.626pt}{0.119pt}}
\multiput(618.00,538.17)(23.700,19.000){2}{\rule{0.313pt}{0.400pt}}
\multiput(643.00,558.58)(0.686,0.495){35}{\rule{0.647pt}{0.119pt}}
\multiput(643.00,557.17)(24.656,19.000){2}{\rule{0.324pt}{0.400pt}}
\multiput(669.00,577.58)(0.785,0.494){29}{\rule{0.725pt}{0.119pt}}
\multiput(669.00,576.17)(23.495,16.000){2}{\rule{0.363pt}{0.400pt}}
\multiput(694.00,593.58)(0.873,0.494){27}{\rule{0.793pt}{0.119pt}}
\multiput(694.00,592.17)(24.353,15.000){2}{\rule{0.397pt}{0.400pt}}
\multiput(720.00,608.58)(0.938,0.494){25}{\rule{0.843pt}{0.119pt}}
\multiput(720.00,607.17)(24.251,14.000){2}{\rule{0.421pt}{0.400pt}}
\multiput(746.00,622.58)(1.052,0.493){23}{\rule{0.931pt}{0.119pt}}
\multiput(746.00,621.17)(25.068,13.000){2}{\rule{0.465pt}{0.400pt}}
\multiput(773.00,635.58)(1.251,0.492){19}{\rule{1.082pt}{0.118pt}}
\multiput(773.00,634.17)(24.755,11.000){2}{\rule{0.541pt}{0.400pt}}
\multiput(800.00,646.59)(1.718,0.489){15}{\rule{1.433pt}{0.118pt}}
\multiput(800.00,645.17)(27.025,9.000){2}{\rule{0.717pt}{0.400pt}}
\multiput(830.00,655.58)(1.642,0.491){17}{\rule{1.380pt}{0.118pt}}
\multiput(830.00,654.17)(29.136,10.000){2}{\rule{0.690pt}{0.400pt}}
\multiput(862.00,665.59)(2.323,0.485){11}{\rule{1.871pt}{0.117pt}}
\multiput(862.00,664.17)(27.116,7.000){2}{\rule{0.936pt}{0.400pt}}
\multiput(893.00,672.59)(2.825,0.477){7}{\rule{2.180pt}{0.115pt}}
\multiput(893.00,671.17)(21.475,5.000){2}{\rule{1.090pt}{0.400pt}}
\multiput(919.00,677.60)(3.990,0.468){5}{\rule{2.900pt}{0.113pt}}
\multiput(919.00,676.17)(21.981,4.000){2}{\rule{1.450pt}{0.400pt}}
\multiput(947.00,681.61)(5.151,0.447){3}{\rule{3.300pt}{0.108pt}}
\multiput(947.00,680.17)(17.151,3.000){2}{\rule{1.650pt}{0.400pt}}
\put(971,684.17){\rule{4.700pt}{0.400pt}}
\multiput(971.00,683.17)(13.245,2.000){2}{\rule{2.350pt}{0.400pt}}
\put(994,686.17){\rule{4.500pt}{0.400pt}}
\multiput(994.00,685.17)(12.660,2.000){2}{\rule{2.250pt}{0.400pt}}
\put(1016,687.67){\rule{4.818pt}{0.400pt}}
\multiput(1016.00,687.17)(10.000,1.000){2}{\rule{2.409pt}{0.400pt}}
\put(1036,688.67){\rule{4.818pt}{0.400pt}}
\multiput(1036.00,688.17)(10.000,1.000){2}{\rule{2.409pt}{0.400pt}}
\put(1056,689.67){\rule{4.577pt}{0.400pt}}
\multiput(1056.00,689.17)(9.500,1.000){2}{\rule{2.289pt}{0.400pt}}
\put(1075,690.67){\rule{4.577pt}{0.400pt}}
\multiput(1075.00,690.17)(9.500,1.000){2}{\rule{2.289pt}{0.400pt}}
\put(1094,691.67){\rule{4.095pt}{0.400pt}}
\multiput(1094.00,691.17)(8.500,1.000){2}{\rule{2.048pt}{0.400pt}}
\put(226.0,163.0){\usebox{\plotpoint}}
\put(1145,692.67){\rule{4.095pt}{0.400pt}}
\multiput(1145.00,692.17)(8.500,1.000){2}{\rule{2.048pt}{0.400pt}}
\put(1111.0,693.0){\rule[-0.200pt]{8.191pt}{0.400pt}}
\put(1209,693.67){\rule{3.614pt}{0.400pt}}
\multiput(1209.00,693.17)(7.500,1.000){2}{\rule{1.807pt}{0.400pt}}
\put(1162.0,694.0){\rule[-0.200pt]{11.322pt}{0.400pt}}
\put(1224.0,695.0){\rule[-0.200pt]{51.071pt}{0.400pt}}
\put(220,695){\raisebox{-.8pt}{\makebox(0,0){$\bullet$}}}
\put(256,648){\raisebox{-.8pt}{\makebox(0,0){$\bullet$}}}
\put(264,625){\raisebox{-.8pt}{\makebox(0,0){$\bullet$}}}
\put(277,592){\raisebox{-.8pt}{\makebox(0,0){$\bullet$}}}
\put(289,564){\raisebox{-.8pt}{\makebox(0,0){$\bullet$}}}
\put(314,532){\raisebox{-.8pt}{\makebox(0,0){$\bullet$}}}
\put(335,495){\raisebox{-.8pt}{\makebox(0,0){$\bullet$}}}
\put(361,453){\raisebox{-.8pt}{\makebox(0,0){$\bullet$}}}
\put(386,417){\raisebox{-.8pt}{\makebox(0,0){$\bullet$}}}
\put(415,400){\raisebox{-.8pt}{\makebox(0,0){$\bullet$}}}
\put(449,392){\raisebox{-.8pt}{\makebox(0,0){$\bullet$}}}
\put(478,388){\raisebox{-.8pt}{\makebox(0,0){$\bullet$}}}
\put(501,395){\raisebox{-.8pt}{\makebox(0,0){$\bullet$}}}
\put(529,407){\raisebox{-.8pt}{\makebox(0,0){$\bullet$}}}
\put(556,425){\raisebox{-.8pt}{\makebox(0,0){$\bullet$}}}
\put(604,473){\raisebox{-.8pt}{\makebox(0,0){$\bullet$}}}
\put(665,542){\raisebox{-.8pt}{\makebox(0,0){$\bullet$}}}
\put(696,580){\raisebox{-.8pt}{\makebox(0,0){$\bullet$}}}
\put(743,622){\raisebox{-.8pt}{\makebox(0,0){$\bullet$}}}
\put(787,653){\raisebox{-.8pt}{\makebox(0,0){$\bullet$}}}
\put(850,673){\raisebox{-.8pt}{\makebox(0,0){$\bullet$}}}
\put(919,683){\raisebox{-.8pt}{\makebox(0,0){$\bullet$}}}
\put(1015,691){\raisebox{-.8pt}{\makebox(0,0){$\bullet$}}}
\put(1416,695){\raisebox{-.8pt}{\makebox(0,0){$\bullet$}}}
\multiput(318.58,853.62)(0.492,-7.131){21}{\rule{0.119pt}{5.633pt}}
\multiput(317.17,865.31)(12.000,-154.308){2}{\rule{0.400pt}{2.817pt}}
\multiput(330.58,695.09)(0.495,-4.753){33}{\rule{0.119pt}{3.833pt}}
\multiput(329.17,703.04)(18.000,-160.044){2}{\rule{0.400pt}{1.917pt}}
\multiput(348.58,531.63)(0.496,-3.351){37}{\rule{0.119pt}{2.740pt}}
\multiput(347.17,537.31)(20.000,-126.313){2}{\rule{0.400pt}{1.370pt}}
\multiput(368.58,402.92)(0.496,-2.340){39}{\rule{0.119pt}{1.948pt}}
\multiput(367.17,406.96)(21.000,-92.958){2}{\rule{0.400pt}{0.974pt}}
\multiput(389.58,308.23)(0.496,-1.631){41}{\rule{0.120pt}{1.391pt}}
\multiput(388.17,311.11)(22.000,-68.113){2}{\rule{0.400pt}{0.695pt}}
\multiput(411.58,238.89)(0.496,-1.122){41}{\rule{0.120pt}{0.991pt}}
\multiput(410.17,240.94)(22.000,-46.943){2}{\rule{0.400pt}{0.495pt}}
\multiput(433.58,191.44)(0.496,-0.646){45}{\rule{0.120pt}{0.617pt}}
\multiput(432.17,192.72)(24.000,-29.720){2}{\rule{0.400pt}{0.308pt}}
\multiput(457.00,161.92)(0.873,-0.494){27}{\rule{0.793pt}{0.119pt}}
\multiput(457.00,162.17)(24.353,-15.000){2}{\rule{0.397pt}{0.400pt}}
\put(483,148.17){\rule{5.700pt}{0.400pt}}
\multiput(483.00,147.17)(16.169,2.000){2}{\rule{2.850pt}{0.400pt}}
\multiput(511.00,150.58)(1.011,0.494){25}{\rule{0.900pt}{0.119pt}}
\multiput(511.00,149.17)(26.132,14.000){2}{\rule{0.450pt}{0.400pt}}
\multiput(539.00,164.58)(0.562,0.496){45}{\rule{0.550pt}{0.120pt}}
\multiput(539.00,163.17)(25.858,24.000){2}{\rule{0.275pt}{0.400pt}}
\multiput(566.58,188.00)(0.497,0.596){49}{\rule{0.120pt}{0.577pt}}
\multiput(565.17,188.00)(26.000,29.803){2}{\rule{0.400pt}{0.288pt}}
\multiput(592.58,219.00)(0.497,0.654){49}{\rule{0.120pt}{0.623pt}}
\multiput(591.17,219.00)(26.000,32.707){2}{\rule{0.400pt}{0.312pt}}
\multiput(618.58,253.00)(0.497,0.713){49}{\rule{0.120pt}{0.669pt}}
\multiput(617.17,253.00)(26.000,35.611){2}{\rule{0.400pt}{0.335pt}}
\multiput(644.58,290.00)(0.497,0.762){47}{\rule{0.120pt}{0.708pt}}
\multiput(643.17,290.00)(25.000,36.531){2}{\rule{0.400pt}{0.354pt}}
\multiput(669.58,328.00)(0.497,0.732){49}{\rule{0.120pt}{0.685pt}}
\multiput(668.17,328.00)(26.000,36.579){2}{\rule{0.400pt}{0.342pt}}
\multiput(695.58,366.00)(0.497,0.742){47}{\rule{0.120pt}{0.692pt}}
\multiput(694.17,366.00)(25.000,35.564){2}{\rule{0.400pt}{0.346pt}}
\multiput(720.58,403.00)(0.497,0.722){47}{\rule{0.120pt}{0.676pt}}
\multiput(719.17,403.00)(25.000,34.597){2}{\rule{0.400pt}{0.338pt}}
\multiput(745.58,439.00)(0.497,0.654){49}{\rule{0.120pt}{0.623pt}}
\multiput(744.17,439.00)(26.000,32.707){2}{\rule{0.400pt}{0.312pt}}
\multiput(771.58,473.00)(0.497,0.620){47}{\rule{0.120pt}{0.596pt}}
\multiput(770.17,473.00)(25.000,29.763){2}{\rule{0.400pt}{0.298pt}}
\multiput(796.58,504.00)(0.497,0.576){49}{\rule{0.120pt}{0.562pt}}
\multiput(795.17,504.00)(26.000,28.834){2}{\rule{0.400pt}{0.281pt}}
\multiput(822.00,534.58)(0.498,0.497){49}{\rule{0.500pt}{0.120pt}}
\multiput(822.00,533.17)(24.962,26.000){2}{\rule{0.250pt}{0.400pt}}
\multiput(848.00,560.58)(0.562,0.496){45}{\rule{0.550pt}{0.120pt}}
\multiput(848.00,559.17)(25.858,24.000){2}{\rule{0.275pt}{0.400pt}}
\multiput(875.00,584.58)(0.637,0.496){41}{\rule{0.609pt}{0.120pt}}
\multiput(875.00,583.17)(26.736,22.000){2}{\rule{0.305pt}{0.400pt}}
\multiput(903.00,606.58)(0.820,0.495){35}{\rule{0.753pt}{0.119pt}}
\multiput(903.00,605.17)(29.438,19.000){2}{\rule{0.376pt}{0.400pt}}
\multiput(934.00,625.58)(0.913,0.494){29}{\rule{0.825pt}{0.119pt}}
\multiput(934.00,624.17)(27.288,16.000){2}{\rule{0.413pt}{0.400pt}}
\multiput(963.00,641.58)(1.156,0.492){19}{\rule{1.009pt}{0.118pt}}
\multiput(963.00,640.17)(22.906,11.000){2}{\rule{0.505pt}{0.400pt}}
\multiput(988.00,652.58)(1.329,0.491){17}{\rule{1.140pt}{0.118pt}}
\multiput(988.00,651.17)(23.634,10.000){2}{\rule{0.570pt}{0.400pt}}
\multiput(1014.00,662.59)(1.789,0.485){11}{\rule{1.471pt}{0.117pt}}
\multiput(1014.00,661.17)(20.946,7.000){2}{\rule{0.736pt}{0.400pt}}
\multiput(1038.00,669.59)(2.380,0.477){7}{\rule{1.860pt}{0.115pt}}
\multiput(1038.00,668.17)(18.139,5.000){2}{\rule{0.930pt}{0.400pt}}
\multiput(1060.00,674.60)(2.967,0.468){5}{\rule{2.200pt}{0.113pt}}
\multiput(1060.00,673.17)(16.434,4.000){2}{\rule{1.100pt}{0.400pt}}
\multiput(1081.00,678.61)(4.034,0.447){3}{\rule{2.633pt}{0.108pt}}
\multiput(1081.00,677.17)(13.534,3.000){2}{\rule{1.317pt}{0.400pt}}
\multiput(1100.00,681.61)(4.034,0.447){3}{\rule{2.633pt}{0.108pt}}
\multiput(1100.00,680.17)(13.534,3.000){2}{\rule{1.317pt}{0.400pt}}
\put(1119,684.17){\rule{3.900pt}{0.400pt}}
\multiput(1119.00,683.17)(10.905,2.000){2}{\rule{1.950pt}{0.400pt}}
\put(1138,686.17){\rule{3.500pt}{0.400pt}}
\multiput(1138.00,685.17)(9.736,2.000){2}{\rule{1.750pt}{0.400pt}}
\put(1155,687.67){\rule{4.336pt}{0.400pt}}
\multiput(1155.00,687.17)(9.000,1.000){2}{\rule{2.168pt}{0.400pt}}
\put(1173,688.67){\rule{3.854pt}{0.400pt}}
\multiput(1173.00,688.17)(8.000,1.000){2}{\rule{1.927pt}{0.400pt}}
\put(1189,689.67){\rule{4.095pt}{0.400pt}}
\multiput(1189.00,689.17)(8.500,1.000){2}{\rule{2.048pt}{0.400pt}}
\put(1206,690.67){\rule{3.614pt}{0.400pt}}
\multiput(1206.00,690.17)(7.500,1.000){2}{\rule{1.807pt}{0.400pt}}
\put(1221,691.67){\rule{3.854pt}{0.400pt}}
\multiput(1221.00,691.17)(8.000,1.000){2}{\rule{1.927pt}{0.400pt}}
\put(1267,692.67){\rule{3.614pt}{0.400pt}}
\multiput(1267.00,692.17)(7.500,1.000){2}{\rule{1.807pt}{0.400pt}}
\put(1237.0,693.0){\rule[-0.200pt]{7.227pt}{0.400pt}}
\put(1338,693.67){\rule{3.132pt}{0.400pt}}
\multiput(1338.00,693.17)(6.500,1.000){2}{\rule{1.566pt}{0.400pt}}
\put(1282.0,694.0){\rule[-0.200pt]{13.490pt}{0.400pt}}
\put(1351.0,695.0){\rule[-0.200pt]{20.476pt}{0.400pt}}
\end{picture}
\vspace{.5 in}
\caption{The scaling form ${\cal G}_{s}(x)$ for the signed vortex correlation
function. At $x=1$ the lower solid curve is the prediction of the
original theory \protect\cite{LIU92b,LIU92a}
that does not treat fluctuations, and
the upper solid curve is the result for the theory presented here which
does include fluctuations.
The dots represent the simulation data \protect\cite{MONDELLO90}.}
\label{FIG:GS}
\end{figure}
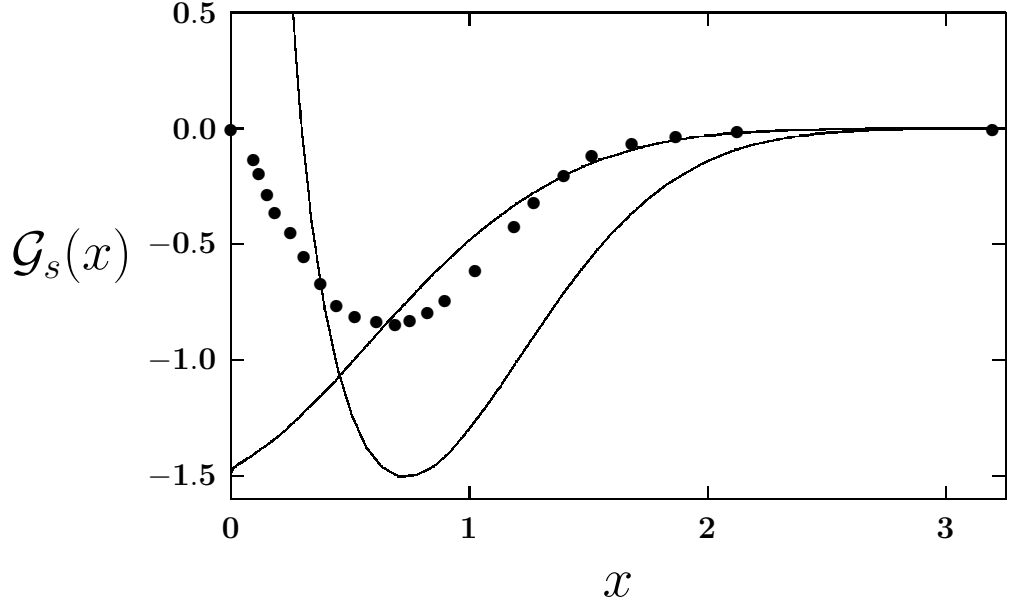
%
% FIGURE 4: UNSIGNED DEFECT-DEFECT CORRELATIONS
%
\begin{figure}
% GNUPLOT: LaTeX picture
\setlength{\unitlength}{0.240900pt}
\ifx\plotpoint\undefined\newsavebox{\plotpoint}\fi
\sbox{\plotpoint}{\rule[-0.200pt]{0.400pt}{0.400pt}}%
\begin{picture}(1500,900)(-150,0)
\font\gnuplot=cmr10 at 10pt
\gnuplot
\sbox{\plotpoint}{\rule[-0.200pt]{0.400pt}{0.400pt}}%
\put(220.0,113.0){\rule[-0.200pt]{0.400pt}{184.048pt}}
\put(220.0,113.0){\rule[-0.200pt]{4.818pt}{0.400pt}}
\put(198,113){\makebox(0,0)[r]{${\bf 0.0}$}}
\put(1416.0,113.0){\rule[-0.200pt]{4.818pt}{0.400pt}}
\put(220.0,352.0){\rule[-0.200pt]{4.818pt}{0.400pt}}
\put(198,352){\makebox(0,0)[r]{${\bf 0.5}$}}
\put(1416.0,352.0){\rule[-0.200pt]{4.818pt}{0.400pt}}
\put(220.0,591.0){\rule[-0.200pt]{4.818pt}{0.400pt}}
\put(198,591){\makebox(0,0)[r]{${\bf 1.0}$}}
\put(1416.0,591.0){\rule[-0.200pt]{4.818pt}{0.400pt}}
\put(220.0,829.0){\rule[-0.200pt]{4.818pt}{0.400pt}}
\put(198,829){\makebox(0,0)[r]{${\bf 1.5}$}}
\put(1416.0,829.0){\rule[-0.200pt]{4.818pt}{0.400pt}}
\put(220.0,113.0){\rule[-0.200pt]{0.400pt}{4.818pt}}
\put(220,68){\makebox(0,0){${\bf 0}$}}
\put(220.0,857.0){\rule[-0.200pt]{0.400pt}{4.818pt}}
\put(594.0,113.0){\rule[-0.200pt]{0.400pt}{4.818pt}}
\put(594,68){\makebox(0,0){${\bf 1}$}}
\put(594.0,857.0){\rule[-0.200pt]{0.400pt}{4.818pt}}
\put(968.0,113.0){\rule[-0.200pt]{0.400pt}{4.818pt}}
\put(968,68){\makebox(0,0){${\bf 2}$}}
\put(968.0,857.0){\rule[-0.200pt]{0.400pt}{4.818pt}}
\put(1342.0,113.0){\rule[-0.200pt]{0.400pt}{4.818pt}}
\put(1342,68){\makebox(0,0){${\bf 3}$}}
\put(1342.0,857.0){\rule[-0.200pt]{0.400pt}{4.818pt}}
\put(220.0,113.0){\rule[-0.200pt]{292.934pt}{0.400pt}}
\put(1436.0,113.0){\rule[-0.200pt]{0.400pt}{184.048pt}}
\put(220.0,877.0){\rule[-0.200pt]{292.934pt}{0.400pt}}
\put(-30,495){\makebox(0,0){\LARGE{${\cal G}_{u}(x)$}}}
\put(828,-22){\makebox(0,0){\LARGE{$x$}}}
\put(220.0,113.0){\rule[-0.200pt]{0.400pt}{184.048pt}}
\put(221,821){\usebox{\plotpoint}}
\put(221.0,818.0){\rule[-0.200pt]{0.400pt}{0.723pt}}
\put(221.0,818.0){\usebox{\plotpoint}}
\put(222,815.67){\rule{0.241pt}{0.400pt}}
\multiput(222.00,816.17)(0.500,-1.000){2}{\rule{0.120pt}{0.400pt}}
\put(222.0,817.0){\usebox{\plotpoint}}
\put(223,813.67){\rule{0.241pt}{0.400pt}}
\multiput(223.00,814.17)(0.500,-1.000){2}{\rule{0.120pt}{0.400pt}}
\put(223.0,815.0){\usebox{\plotpoint}}
\put(224,814){\usebox{\plotpoint}}
\put(224,812.67){\rule{0.241pt}{0.400pt}}
\multiput(224.00,813.17)(0.500,-1.000){2}{\rule{0.120pt}{0.400pt}}
\put(225,811.67){\rule{0.241pt}{0.400pt}}
\multiput(225.00,812.17)(0.500,-1.000){2}{\rule{0.120pt}{0.400pt}}
\put(226,810.67){\rule{0.241pt}{0.400pt}}
\multiput(226.00,811.17)(0.500,-1.000){2}{\rule{0.120pt}{0.400pt}}
\put(227,809.67){\rule{0.241pt}{0.400pt}}
\multiput(227.00,810.17)(0.500,-1.000){2}{\rule{0.120pt}{0.400pt}}
\put(228,808.67){\rule{0.241pt}{0.400pt}}
\multiput(228.00,809.17)(0.500,-1.000){2}{\rule{0.120pt}{0.400pt}}
\put(229,807.67){\rule{0.241pt}{0.400pt}}
\multiput(229.00,808.17)(0.500,-1.000){2}{\rule{0.120pt}{0.400pt}}
\put(230,806.67){\rule{0.241pt}{0.400pt}}
\multiput(230.00,807.17)(0.500,-1.000){2}{\rule{0.120pt}{0.400pt}}
\put(231,805.17){\rule{0.482pt}{0.400pt}}
\multiput(231.00,806.17)(1.000,-2.000){2}{\rule{0.241pt}{0.400pt}}
\put(233,803.67){\rule{0.482pt}{0.400pt}}
\multiput(233.00,804.17)(1.000,-1.000){2}{\rule{0.241pt}{0.400pt}}
\put(235,802.17){\rule{0.482pt}{0.400pt}}
\multiput(235.00,803.17)(1.000,-2.000){2}{\rule{0.241pt}{0.400pt}}
\put(237,800.17){\rule{0.700pt}{0.400pt}}
\multiput(237.00,801.17)(1.547,-2.000){2}{\rule{0.350pt}{0.400pt}}
\multiput(240.00,798.95)(0.462,-0.447){3}{\rule{0.500pt}{0.108pt}}
\multiput(240.00,799.17)(1.962,-3.000){2}{\rule{0.250pt}{0.400pt}}
\multiput(243.00,795.95)(0.685,-0.447){3}{\rule{0.633pt}{0.108pt}}
\multiput(243.00,796.17)(2.685,-3.000){2}{\rule{0.317pt}{0.400pt}}
\multiput(247.00,792.94)(0.627,-0.468){5}{\rule{0.600pt}{0.113pt}}
\multiput(247.00,793.17)(3.755,-4.000){2}{\rule{0.300pt}{0.400pt}}
\multiput(252.00,788.93)(0.581,-0.482){9}{\rule{0.567pt}{0.116pt}}
\multiput(252.00,789.17)(5.824,-6.000){2}{\rule{0.283pt}{0.400pt}}
\multiput(259.00,782.93)(0.569,-0.485){11}{\rule{0.557pt}{0.117pt}}
\multiput(259.00,783.17)(6.844,-7.000){2}{\rule{0.279pt}{0.400pt}}
\multiput(267.00,775.93)(0.671,-0.482){9}{\rule{0.633pt}{0.116pt}}
\multiput(267.00,776.17)(6.685,-6.000){2}{\rule{0.317pt}{0.400pt}}
\multiput(275.00,769.93)(0.560,-0.488){13}{\rule{0.550pt}{0.117pt}}
\multiput(275.00,770.17)(7.858,-8.000){2}{\rule{0.275pt}{0.400pt}}
\multiput(284.00,761.93)(0.560,-0.488){13}{\rule{0.550pt}{0.117pt}}
\multiput(284.00,762.17)(7.858,-8.000){2}{\rule{0.275pt}{0.400pt}}
\multiput(293.00,753.93)(0.611,-0.489){15}{\rule{0.589pt}{0.118pt}}
\multiput(293.00,754.17)(9.778,-9.000){2}{\rule{0.294pt}{0.400pt}}
\multiput(304.00,744.92)(0.496,-0.492){19}{\rule{0.500pt}{0.118pt}}
\multiput(304.00,745.17)(9.962,-11.000){2}{\rule{0.250pt}{0.400pt}}
\multiput(315.00,733.92)(0.582,-0.492){21}{\rule{0.567pt}{0.119pt}}
\multiput(315.00,734.17)(12.824,-12.000){2}{\rule{0.283pt}{0.400pt}}
\multiput(329.00,721.92)(0.534,-0.494){25}{\rule{0.529pt}{0.119pt}}
\multiput(329.00,722.17)(13.903,-14.000){2}{\rule{0.264pt}{0.400pt}}
\multiput(344.00,707.92)(0.558,-0.495){31}{\rule{0.547pt}{0.119pt}}
\multiput(344.00,708.17)(17.865,-17.000){2}{\rule{0.274pt}{0.400pt}}
\multiput(363.00,690.92)(0.611,-0.495){33}{\rule{0.589pt}{0.119pt}}
\multiput(363.00,691.17)(20.778,-18.000){2}{\rule{0.294pt}{0.400pt}}
\multiput(385.00,672.92)(0.651,-0.496){37}{\rule{0.620pt}{0.119pt}}
\multiput(385.00,673.17)(24.713,-20.000){2}{\rule{0.310pt}{0.400pt}}
\multiput(411.00,652.92)(0.753,-0.495){33}{\rule{0.700pt}{0.119pt}}
\multiput(411.00,653.17)(25.547,-18.000){2}{\rule{0.350pt}{0.400pt}}
\multiput(438.00,634.92)(0.881,-0.494){29}{\rule{0.800pt}{0.119pt}}
\multiput(438.00,635.17)(26.340,-16.000){2}{\rule{0.400pt}{0.400pt}}
\multiput(466.00,618.92)(1.052,-0.493){23}{\rule{0.931pt}{0.119pt}}
\multiput(466.00,619.17)(25.068,-13.000){2}{\rule{0.465pt}{0.400pt}}
\multiput(493.00,605.92)(1.277,-0.491){17}{\rule{1.100pt}{0.118pt}}
\multiput(493.00,606.17)(22.717,-10.000){2}{\rule{0.550pt}{0.400pt}}
\multiput(518.00,595.93)(1.865,-0.485){11}{\rule{1.529pt}{0.117pt}}
\multiput(518.00,596.17)(21.827,-7.000){2}{\rule{0.764pt}{0.400pt}}
\multiput(543.00,588.93)(2.714,-0.477){7}{\rule{2.100pt}{0.115pt}}
\multiput(543.00,589.17)(20.641,-5.000){2}{\rule{1.050pt}{0.400pt}}
\multiput(568.00,583.94)(3.552,-0.468){5}{\rule{2.600pt}{0.113pt}}
\multiput(568.00,584.17)(19.604,-4.000){2}{\rule{1.300pt}{0.400pt}}
\multiput(593.00,579.95)(5.374,-0.447){3}{\rule{3.433pt}{0.108pt}}
\multiput(593.00,580.17)(17.874,-3.000){2}{\rule{1.717pt}{0.400pt}}
\put(618,576.67){\rule{6.023pt}{0.400pt}}
\multiput(618.00,577.17)(12.500,-1.000){2}{\rule{3.011pt}{0.400pt}}
\put(694,576.67){\rule{6.263pt}{0.400pt}}
\multiput(694.00,576.17)(13.000,1.000){2}{\rule{3.132pt}{0.400pt}}
\put(720,578.17){\rule{5.300pt}{0.400pt}}
\multiput(720.00,577.17)(15.000,2.000){2}{\rule{2.650pt}{0.400pt}}
\put(746,579.67){\rule{6.504pt}{0.400pt}}
\multiput(746.00,579.17)(13.500,1.000){2}{\rule{3.252pt}{0.400pt}}
\put(773,581.17){\rule{5.500pt}{0.400pt}}
\multiput(773.00,580.17)(15.584,2.000){2}{\rule{2.750pt}{0.400pt}}
\put(800,583.17){\rule{6.100pt}{0.400pt}}
\multiput(800.00,582.17)(17.339,2.000){2}{\rule{3.050pt}{0.400pt}}
\put(830,584.67){\rule{7.709pt}{0.400pt}}
\multiput(830.00,584.17)(16.000,1.000){2}{\rule{3.854pt}{0.400pt}}
\put(862,586.17){\rule{6.300pt}{0.400pt}}
\multiput(862.00,585.17)(17.924,2.000){2}{\rule{3.150pt}{0.400pt}}
\put(643.0,577.0){\rule[-0.200pt]{12.286pt}{0.400pt}}
\put(919,587.67){\rule{6.745pt}{0.400pt}}
\multiput(919.00,587.17)(14.000,1.000){2}{\rule{3.373pt}{0.400pt}}
\put(947,588.67){\rule{5.782pt}{0.400pt}}
\multiput(947.00,588.17)(12.000,1.000){2}{\rule{2.891pt}{0.400pt}}
\put(893.0,588.0){\rule[-0.200pt]{6.263pt}{0.400pt}}
\put(1016,589.67){\rule{4.818pt}{0.400pt}}
\multiput(1016.00,589.17)(10.000,1.000){2}{\rule{2.409pt}{0.400pt}}
\put(971.0,590.0){\rule[-0.200pt]{10.840pt}{0.400pt}}
\put(1036.0,591.0){\rule[-0.200pt]{96.360pt}{0.400pt}}
\put(220,113){\raisebox{-.8pt}{\makebox(0,0){$\bullet$}}}
\put(256,174){\raisebox{-.8pt}{\makebox(0,0){$\bullet$}}}
\put(264,205){\raisebox{-.8pt}{\makebox(0,0){$\bullet$}}}
\put(277,248){\raisebox{-.8pt}{\makebox(0,0){$\bullet$}}}
\put(289,284){\raisebox{-.8pt}{\makebox(0,0){$\bullet$}}}
\put(314,327){\raisebox{-.8pt}{\makebox(0,0){$\bullet$}}}
\put(335,376){\raisebox{-.8pt}{\makebox(0,0){$\bullet$}}}
\put(361,431){\raisebox{-.8pt}{\makebox(0,0){$\bullet$}}}
\put(386,482){\raisebox{-.8pt}{\makebox(0,0){$\bullet$}}}
\put(415,509){\raisebox{-.8pt}{\makebox(0,0){$\bullet$}}}
\put(449,523){\raisebox{-.8pt}{\makebox(0,0){$\bullet$}}}
\put(478,531){\raisebox{-.8pt}{\makebox(0,0){$\bullet$}}}
\put(501,530){\raisebox{-.8pt}{\makebox(0,0){$\bullet$}}}
\put(529,530){\raisebox{-.8pt}{\makebox(0,0){$\bullet$}}}
\put(556,523){\raisebox{-.8pt}{\makebox(0,0){$\bullet$}}}
\put(604,520){\raisebox{-.8pt}{\makebox(0,0){$\bullet$}}}
\put(665,537){\raisebox{-.8pt}{\makebox(0,0){$\bullet$}}}
\put(696,549){\raisebox{-.8pt}{\makebox(0,0){$\bullet$}}}
\put(743,568){\raisebox{-.8pt}{\makebox(0,0){$\bullet$}}}
\put(787,584){\raisebox{-.8pt}{\makebox(0,0){$\bullet$}}}
\put(850,585){\raisebox{-.8pt}{\makebox(0,0){$\bullet$}}}
\put(919,587){\raisebox{-.8pt}{\makebox(0,0){$\bullet$}}}
\put(1015,585){\raisebox{-.8pt}{\makebox(0,0){$\bullet$}}}
\put(1416,591){\raisebox{-.8pt}{\makebox(0,0){$\bullet$}}}
\end{picture}
\vspace{.5 in}
\caption{The scaling form  ${\cal G}_{u}(x)$ for the unsigned vortex
correlation function. The solid curve is the result for the theory
presented here. The dots represent the simulation data
\protect\cite{MONDELLO90}.}
\label{FIG:GU}
\end{figure}
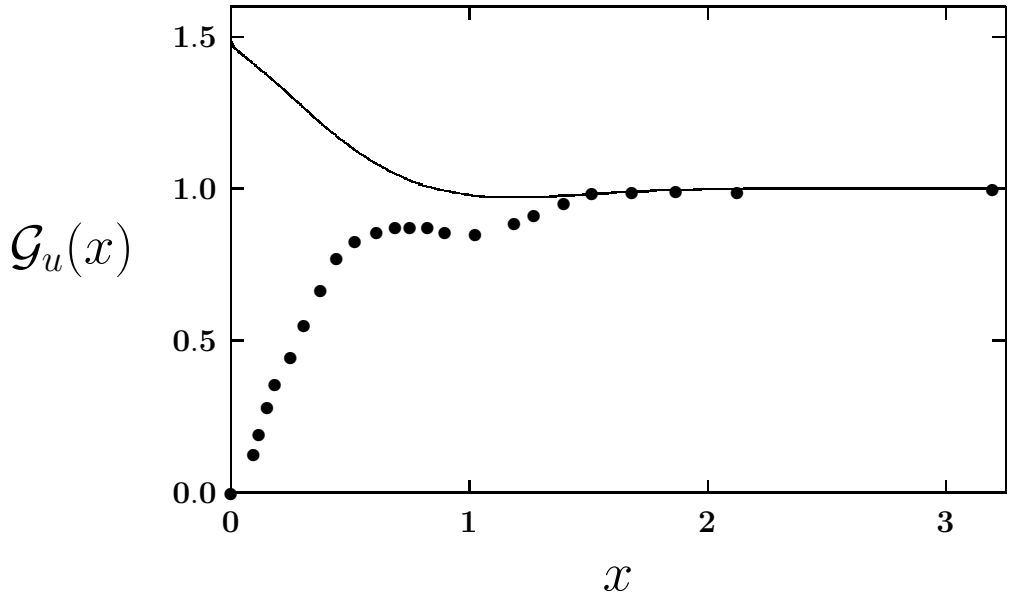

\begin{thebibliography}{99}
\bibitem{NEUTRALITY} We assume overall topological charge neutrality so
                     $\langle \rho (1) \rangle = \sum_{\alpha} q_{\alpha} = 0$.
\bibitem{NOTATION} The factor of $(8 \mu/S^{(2)})^{n} \Gamma^{2}(1 + n/2)/
                   (n!)^{2}$, not present in earlier
                   formulae for ${\cal G}_{s}$ \cite{LIU92b,MAZENKO96},
                   appears in
                   (\ref{EQ:DEFECTDEFECTCOR}) because in this paper the
                   time-dependent amplitude relating the
                   unscaled quantity $\tilde{G}_{s}$ to the scaled
		   quantity ${\cal G}_{s}$ is
                   $[n_{0}(t)]^{2}$ instead of $L^{-2n}$.
\bibitem{NORMALIZATION} Previously, when comparing theory and simulation for
                        the signed vortex correlation function
		        \cite{LIU92b,MAZENKO96} appearing in Figure
                        \ref{FIG:GS}, the amplitude of the
                        simulation curve was a second free fitting parameter.
                        Here, the relation between the theoretical and
                        simulation amplitudes is fixed because the defect
                        density is used to normalize the curves.
\bibitem{CORE} An additional consideration is that the vortex cores in the
               simulation are found to be eight lattice sites wide, a distance
               comparable to the position of the peak in figure 9 of
               \cite{MONDELLO90}.
\bibitem{CAVEAT} One caveat is that the results of Appendix C follow from the
                 small-$x$ expansion (\ref{EQ:FEXPANSION}) for $f$.
                 In \cite{MAZENKO96} the proper treatment of fluctuations
                 allowed us to use this expansion for $n=d=2$. We expect to
                 be able to extend \cite{MAZENKO96}
                 and construct $f$ to be analytic to ${\cal O}(x^{4})$ for
                 $n>2$.
\end{thebibliography}
\end{document}